\DocumentMetadata{%
	pdfstandard=A-3u,% A-2b, A-2u, A-3b, or A-3u
	pdfversion=1.7,
	lang=en-US,
}

\documentclass[balance,nocopyright,upint,varvw,hyphenate,barcolor=Goldenrod3,german]{asmejour}% 

\allowdisplaybreaks % from amsmath package, allows multiline equations to break across pages (delete if not wanted)
\usepackage{hologo} % access various latex logos if needed
\usepackage{comment} % for comment environment

\usepackage[figuresright]{rotating} % sidewaystable*, sidewaysfigure*
\usepackage{booktabs} % \toprule, \midrule, \bottomrule
\usepackage{array}    % p{..} column types

\usepackage{makecell}   % makecell
\usepackage{placeins}   % \FloatBarrier

\newcommand\blfootnote[1]{%
  \begingroup
  \renewcommand\thefootnote{}% 
  \footnote{#1}%
  \addtocounter{footnote}{-1}% footnote counter
  \endgroup
}

\hypersetup{%
	pdfauthor={Suk Ki Lee},                       		   	% <=== change to YOUR name[s]!
	pdftitle={Generative Model Predictive Control in Manufacturing Processes: A Review},                  	% <=== change to YOUR pdf file title
	pdfkeywords={Generative Model Predictive Control in Manufacturing Processes: A Review},% <=== change to YOUR pdf keywords
	pdfsubject = {},			% <=== change to YOUR subject
}

\begin{document}

% Change to your author name[s] and addresses, in the desired order of authors.
% First name, middle initial, last name
% Use title case (upper and lower case letters)
% Note usage below for corresponding author.
\SetAuthorBlock{Suk Ki Lee}{School of Manufacturing Systems and Networks,\\
   Arizona State University,\\
   Mesa, AZ, USA\\
   email: sukki.lee@asu.edu
}

\SetAuthorBlock{Ronnie F. P. Stone}{Walker Department of Mechanical Engineering,\\
University of Texas at Austin,\\
Austin, TX, USA\\
e-mail: ronnie.stone@utexas.edu
} 

\SetAuthorBlock{Max Gao}{School of Manufacturing Systems and Networks,\\
   Arizona State University,\\
   Meas, AZ, USA\\
   email: mgao18@asu.edu
} 

\SetAuthorBlock{Wenlong Zhang}{School of Manufacturing Systems and Networks,\\
    Arizona State University,\\
    Mesa, AZ, USA\\	
    e-mail: Wenlong.Zhang@asu.edu
} 

\SetAuthorBlock{Zhenghui Sha}{Walker Department of Mechanical Engineering,\\
University of Texas at Austin,\\
Austin, TX, USA\\\
e-mail: zsha@austin.utexas.edu
} 

\SetAuthorBlock{Hyunwoong Ko\CorrespondingAuthor}{%
    School of Manufacturing Systems and Networks,\\
    Arizona State University,\\
    Mesa, AZ, USA\\	
    email: hyunwoong.ko@asu.edu
}

% To label one or more corresponding authors put "Name\CorrespondingAuthor". No space after "Name".
% An optional argument can be added if email is not in address block as
%      "Name\CorrespondingAuthor{write@to.me}"
% Can also include multiple emails and use the command more than once for multiple corresponding authors,
%      "Name\CorrespondingAuthor{write@to.him, write@to.her}"

\begin{comment}
\SetAuthorBlock{Author Name[s]}{Department of Mechanical Engineering,\\
   Institution or Company Name,\\
   Street address,\\
   City, State, Country\\
   email: xxx@yyy.zzz
} 
\end{comment}

%%% Change to your paper title. Can insert line breaks if you wish (otherwise breaks are selected automatically).
\title{Generative Model Predictive Control in Manufacturing Processes: A Review}

%%% Change these to your keywords.  Keywords are automatically printed at the end of the abstract.
%%% This command must come BEFORE the end of the abstract.
%%% If you don't want keywords, omit the \keyword{..} command.

%% Abstract should be no more than 250 words
\begin{abstract}
Manufacturing processes are inherently dynamic and uncertain, with varying parameters and nonlinear behaviors, making robust control essential for maintaining quality and reliability.
Traditional control methods often fail under these conditions due to their reactive nature. 
Model Predictive Control (MPC) has emerged as a more advanced framework, leveraging process models to predict future states and optimize control actions. 
However, MPC relies on simplified models that often fail to capture complex dynamics, and it struggles with accurate state estimation and handling the propagation of uncertainty in manufacturing environments.
Machine learning (ML) has been introduced to enhance MPC by modeling nonlinear dynamics and learning latent representations that support predictive modeling, state estimation, and optimization. 
Yet existing ML-driven MPC approaches remain deterministic and correlation-focused, motivating the exploration of generative.
Generative ML offers new opportunities by learning data distributions, capturing hidden patterns, and inherently managing uncertainty, thereby complementing MPC. 
This review highlights five representative methods and examines how each has been integrated into MPC components, including predictive modeling, state estimation, and optimization. 
By synthesizing these cases, we outline the common ways generative ML can systematically enhance MPC and provide a framework for understanding its potential in diverse manufacturing processes.
We identify key research gaps, propose future directions, and use a representative case to illustrate how generative ML–driven MPC can extend broadly across manufacturing.
Taken together, this review positions generative ML not as an incremental add-on but as a transformative approach to reshape predictive control for next-generation manufacturing systems.

\end{abstract}

\date{Version \versionno, \today}%% You can modify this information as desired. 
							%% Putting \date{} will suppress any date.  
							%% If this command is omitted, date defaults to \today
							%% This command must come somewhere before \maketitle

\maketitle %% This command creates the author/title/abstract block. Essential!

\blfootnote{This paper builds on a work previously presented at the ASME 2025 International Design Engineering Technical Conferences \& Computers and Information in Engineering Conference (IDETC/CIE), Paper No.\ DETC2025-169728 \cite{lee2025generative}, with revisions and extensions for journal submission.}
%%%%%%%%%%%%%%%%%%%%%%%%%%%%%%%%%%%%%%%%%%%%%%%%%%%%%%%%%%%%%%%%%%%%%%%%%%%%%%%%%%%%%%%%%%%%%%%%%%%%%%%
%%%%%%%%%%%%%%%%%%%%%  End of fields to be completed. Now write! %%%%%%%%%%%%%%%%%%%%%%%%%%%%%%%%%%%%%%

\section{Introduction}
% (Explain dynamic characteristics and uncertainties of manufacturing processes.)
Manufacturing is becoming increasingly complex and dynamic, driven by the demand for products with higher complexity, quality improvement, process efficiency, and manufacturing flexibility \cite{qu2019smart, tao2018data}. 
These changes in manufacturing environments evoke unprecedented challenges in handling processes with rapid parameter variations, unpredictable process dynamics, and uncertainties \cite{peres2020industrial, jan2023artificial}.
% Dynamic Characteristics
During manufacturing processes, rapid and ongoing changes occur in different subsets of control parameters over time, with some parameters remaining stable while others vary. 
These variations in parameters influence system behavior and outputs, and the resulting responses often manifest in nonlinear ways. 
%In this paper, we use the term `nonlinear` to refer to any response that departs from a linear relationship, even if not formally demonstrated. 
%Together, these aspects constitute the dynamic characteristics of manufacturing processes. \textcolor{blue}{RS: I think that it is not clear what the message here is. Are you trying to give a definition of dynamic characteristics? Is it being defined here as a combination of parametric variation and nonlinearity?}

% Uncertainties
Along with these dynamic characteristics, manufacturing processes also exhibit various forms of uncertainty.
Uncertainty arises from the inherent variability of materials, environmental conditions, and process execution \cite{mahadevan2022uncertainty}. 
Those uncertainties can be classified into epistemic and aleatoric types \cite{mahadevan2022uncertainty, nannapaneni2014uncertainty}. 
Epistemic uncertainty stems from incomplete knowledge or modeling simplifications induced by assumptions and approximations, while aleatoric uncertainty arises from inherent randomness in materials and processes \cite{mahadevan2022uncertainty, nannapaneni2014uncertainty}. 
Such uncertainty is often expressed through unpredictable dynamics, where process responses deviate from expected behavior even under the same input conditions. Therefore, uncertainty in manufacturing arises not only from gaps in understanding the underlying process physics but also from the inherent variability in the expected output due to randomness. Each of these factors introduces significant challenges for effective process control.

% (Illustrate dynamic characteristics and uncertainties of single process examples.)
% Ex) AM, In-space Manufacturing, Single Process of Semiconductor Manufacturing
These dynamic characteristics and uncertainties appear in various manufacturing scenarios.
Representative examples of single-process and multi-process manufacturing are summarized in Table~\ref{tab:MFG_process_parameter_outputs}, along with their control parameters and system outputs, which implicitly reflect sources of uncertainty.
Cases where a single transformation mechanism dominates the operation can be regarded as a single-process operation, such as laser powder bed fusion (LPBF), direct energy deposition (DED), wire arc additive manufacturing, or a unit process of semiconductor manufacturing, including deposition, etching, and chemical–mechanical polishing \cite{mahadevan2022uncertainty, espadinha2021review, qin2006semiconductor,lee2024amtransformer,han2025physics}. 
For example, variations of process parameters in laser-based additive manufacturing (AM), such as laser power, scanning speed, melting location, and material properties, exhibit ongoing variation.
These variations lead to complex interactions that challenge control systems \cite{everton2016review, cai2023review}. 
Often, these variations result in nonlinear behaviors, where minor changes in input parameters can disproportionately affect melt pool geometry and material properties, as evidenced by thermographic and high-speed imaging studies \cite{zhang2018extraction}. 
Process uncertainties, such as inconsistencies in material properties or environmental factors, further complicate manufacturing control \cite{mahadevan2022uncertainty}. 
In laser-based processes, phenomena such as plume and spatter formation introduce unpredictable melt pool behavior \cite{zhang2018extraction}.

% (Illustrate dynamic characteristics and uncertainties of multi-process examples. (Examples show more complex, dynamic, and higher uncertainty))
% EX) Multi-robot Manufacturing, Whole Process Semiconductor Manufacturing (Entire Fab)
In more complex cases involving the coordination of multiple transformation mechanisms, these are called as multi-process or integrated systems, where mechanisms interact either sequentially or in parallel. 
%\textcolor{blue}{RS: We may want to be careful with our definitions. We are currently defining semiconductor steps (e.g., ion implantation, etching) as single-process. But then we define multi-process systems as sequential steps that may need to account for intra-process interactions. However, this is exactly what happens, and thus needs to be accounted for, in semiconductor foundries. You actually mention this below, so maybe the logic needs to be adjusted.}
Such systems also exhibit dynamic characteristics and uncertainties, but the nature of these factors differs from those in single-process operations. 
In multi-process settings, it is necessary to account for inter-process interactions and the propagation of uncertainties across processes, in addition to the intra-process dynamics typically observed in single-process operations. 
Representative examples include multi-robot manufacturing, where multiple robots coordinate tasks in parallel under shared objectives, and semiconductor fabrication, where multiple unit processes are integrated into a sequential production flow \cite{espadinha2021review, qin2006semiconductor, arai2002advances, parker2016multiple, rizk2019cooperative}.

% Multi-robot example
The multi-robot AM application exemplifies the dynamic characteristics and uncertainties of multi-process systems. 
The presence of multiple robots introduces several additional steps, such as division of labor, task scheduling, and motion planning. 
% Parameter Changes
Relevant parameters include the number of robots, their positions and kinematic properties, as well as printing-related parameters such as nozzle velocity, deposition rate, and thermal profiles, all of which may fluctuate over time or differ across robots.
Such changes escalate process complexity through requirements for precise synchronization, dynamic path planning, and coordinated behavior control \cite{zhang2022aerial}.
% Nonlinear behavior
Multiple robots often lead to nonlinear system behavior. 
For example, the presence of an additional robot in the shared workspace introduces new complexities in partitioning, scheduling, and path planning, so the expected linear reduction in build time is rarely achieved \cite{stone2025safezone, silver2005cooperative, poudel2020heuristic}. 
% Uncertainties
Uncertainties typically manifest as pose errors (e.g., position uncertainty) during motion planning as robots traverse the workspace \cite{stone2025risk, elagandula2020multi}. 
Beyond these motion-related uncertainties, the division of labor introduces additional uncertainties at the interfaces between deposition regions, which affect the mechanical integrity of the build \cite{mensch2024real}.
Slight interfacial mismatches in deposition conditions between robots can further lead to uneven layer thickness at the interface regions, potentially causing nozzle-to-part collisions \cite{mensch2024real}.

In addition to the dynamics and uncertainties related to processes, environmental factors introduce further layers of uncertainty that become more severe under extreme conditions, for example in-space manufacturing.
Manufacturing in space introduces unique complexities due to microgravity conditions that fundamentally alter material flow behaviors, thermal gradients, and solidification mechanisms \cite{sacco2019additive, kalaycioglu2023passivity}. 
Additionally, extreme space conditions, including radiation exposure, vacuum, and temperature fluctuations, significantly impact material properties and process stability \cite{zocca2022challenges, kalaycioglu2023passivity}.
Similar challenges are also expected to arise in other extreme environments, where harsh conditions further amplify process dynamics and uncertainty.

%%%% TABLE %%%%%

\begin{sidewaystable*}[p]
  \centering
  \small % size
  \setlength{\tabcolsep}{5pt}
  \renewcommand{\arraystretch}{1.5}
  \caption{Representative Manufacturing Processes and Their Control Parameters and System Outputs}
  \label{tab:MFG_process_parameter_outputs}
  \begin{tabular}{
    >{\raggedright\arraybackslash}p{2.0cm}  % Scope
    >{\raggedright\arraybackslash}p{2.8cm}  % Category
    >{\raggedright\arraybackslash}p{2.8cm}  % Application Type
    >{\raggedright\arraybackslash}p{6.0cm}  % Control Parameters
    >{\raggedright\arraybackslash}p{6.0cm}  % System Outputs
    >{\raggedright\arraybackslash}p{2.0cm}  % Representative References
  }
  \toprule
  \textbf{Scope} & \textbf{Category} & \textbf{Application Type} & \textbf{Control Parameters} & \textbf{System Outputs} & \textbf{Representative References} \\
  \midrule

  Single-process & Additive Manufacturing (AM) & LPBF &
  Laser power, scan speed, layer thickness, beam offset, hatch spacing, toolpath, chamber temperature &
  Melt pool size, surface roughness, porosity, microstructure, residual stress, mechanical properties &
  \cite{ko2023framework, gunasegaram2024machine, lee2024amtransformer,knaak2021improving} \\

   &                        & DED &
  Laser/arc power, travel speed, powder/wire feed rate, deposition path, overlap ratio, shielding gas, substrate preheat &
  Bead width/height, porosity, cracks, microstructure, residual stress, mechanical properties &
  \cite{svetlizky2021directed, ahn2021directed} \\

   &                        & WAAM &
  Current, voltage, arc length, torch speed, wire feed speed, shielding gas, inter-pass temperature, deposition path &
  Bead geometry, waviness, porosity, microstructure, residual stress, mechanical properties &
  \cite{wu2018review, treutler2021current} \\

   &                        & AJP &
  Atomizer flow, sheath gas flow, printing speed, nozzle-substrate distance, substrate temperature &
  Line width, line height, porosity, conductivity, adhesion &
  \cite{elhambakhsh2025generative, guo2024applications} \\

   & Semiconductor & \makecell[l]{Deposition \\ (CVD, ALD, PVD)} &
  Gas/precursor flow rate, chamber pressure, substrate temperature, deposition time/ALD cycles, plasma/RF power &
  Film thickness, uniformity, composition/stoichiometry, defect density &
  \cite{han2025physics, sheng2018review, tang2024optimizing, kimaev2019nonlinear} \\

   &  & Etching \newline (Dry Etching, RIE, ICP-RIE, ALE) &
  Etchant gas flow/composition, RF/ICP power, chamber pressure, bias voltage, substrate temperature, etch time/ALE cycles &
  Etch rate, selectivity, anisotropy (sidewall profile), etch depth, uniformity, surface damage &
  \cite{racka2021review, huff2021recent} \\

  Multi-process & Multi-robot Manufacturing & AM &
  Number of robots, layer geometry, printer locations, toolpath, scheduling approach &
  Processing time, part strength &
  \cite{stone2023print, stone2025safezone, ebert2025noodleprint, krishnamurthy2022layerlock, poudel2023decentralized} \\

   &                        & Assembly &
  Robot kinematics, mobile base dynamics, velocity constraints, number of robots, cell layout &
  Assembly force profile, global success, processing time &
  \cite{stroupe2005behavior, knepper2013ikeabot, papakostas2011industrial} \\

   &                        & Transport &
  Robot initial and final positions, number of robots, job size and locations &
  Processing time, global success &
  \cite{stone2025risk, elagandula2020multi} \\
  \bottomrule
  \end{tabular}
\end{sidewaystable*}

%%%%%%%%%%%%%%%%

% (Explain why the manufacturing process cannot be properly handled with existing control approaches. Introduce MPC and its limitations.)
% (This paragraph explains the challenges that show the reasons why we need MPC for manufacturing.)
Addressing these challenges requires advanced control strategies that can mitigate defects and improve manufacturing outcomes \cite{cai2023review, gunasegaram2024machine}. 
Traditional control methods, such as proportional–integral-derivative (PID) or rule-based approaches, often rely on simplified assumptions, making them unreliable under the dynamic characteristics and uncertainties of modern manufacturing \cite{Schwenzer21, mayne2014model}. 
Model Predictive Control (MPC) has emerged as a powerful framework because it explicitly uses a given process model to predict future states and optimize control actions \cite{Schwenzer21, mayne2014model}. 
However, MPC depends heavily on accurate models and state estimation. 
When the process is highly nonlinear or high-dimensional, building such models is difficult, and neglecting uncertainties can compromise control performance.

% (Introduce ML and ML-driven MPC with their limitations. Introduce Generative ML and Generative ML-driven MPC.)
To overcome the limitations of MPC, machine learning (ML) has been introduced to enhance its performance. 
ML models can approximate complex, nonlinear dynamics through data-driven approaches and support different components of MPC. 
%\textcolor{blue}{RS: I think it would be best to succinctly explain here what exactly is ML doing for MPC. I think "enhance performance" and "support different components" is a bit too generic. For example, we just mentioned that MPC struggles with model accuracy, particularly in nonlinear or high-dimensional cases. How then, exactly, is ML bridging this gap?}
ML-driven MPC has shown potential in various manufacturing scenarios, yet it still faces challenges such as over-reliance on training data and limited capability in handling uncertainties. 
To address these issues, generative ML offers new possibilities. 
Unlike conventional ML, generative approaches provide a stronger ability to understand data distributions and manage uncertainties. 
As a result, integrating generative models with MPC could enable more proactive and reliable control in manufacturing processes. 
%\textcolor{blue}{RS: Same comment here for generative ML-driven MPC. I do not think we made our case strong enough here. Naturally, the reader will get a better understanding from Sections 2, 3, and 4, but I think that the details are too vague in the introduction.}

% (Summary of the study addressing the challenge)
Although generative ML shows unprecedented potential for enhancing MPC, its integration in manufacturing is still at a very early stage. 
Few studies have attempted to link generative ML methods with MPC, and their potential has not been fully understood or adapted to manufacturing contexts. 
This review seeks to provide guidance by clarifying the principles and capabilities of generative ML and by exploring how these can be systematically connected to the principles of MPC. 
Through this effort, we aim to offer insights that can support future research on generative ML–driven MPC and its applications in manufacturing.

% (Remainder of the paper.)
The remainder of this paper is structured as follows.
Section 2 introduces MPC, its key principles, and main challenges.
Section 3 reviews ML-driven MPC, outlining the core capabilities of ML, how they have been integrated into MPC components, and the key challenges that remain.
Section 4 introduces generative ML, explains its capabilities, and reviews key methods. 
It then examines how these methods can be combined with MPC and what possibilities they open for manufacturing control.
Section 5 identifies current research gaps, while Section 6 discusses future research directions and illustrates a representative manufacturing case to show the potential of generative ML–driven MPC.
Finally, Section 7 provides concluding remarks.

%%%%%%%%%%%%%%%%%%%%%%%%%%%%%%%%%%%%%%%%%%%%%%%%%%%%%%%%%%%%%%%%%%%%%%
\section{Model Predictive Control}
To address the many control challenges presented in manufacturing, more intelligent strategies are required. MPC represents a versatile solution ideal for settings demanding high reliability despite system complexity. 
This section introduces the definition and primary attributes of MPC, and discusses its key challenges facing the control strategy. 
% Section opening paragraph here
% Note: Resolved minor suggested formatting edits
\subsection{Definition}
\begin{figure}[htbp]
    \centering
    \includegraphics[width=1\linewidth]{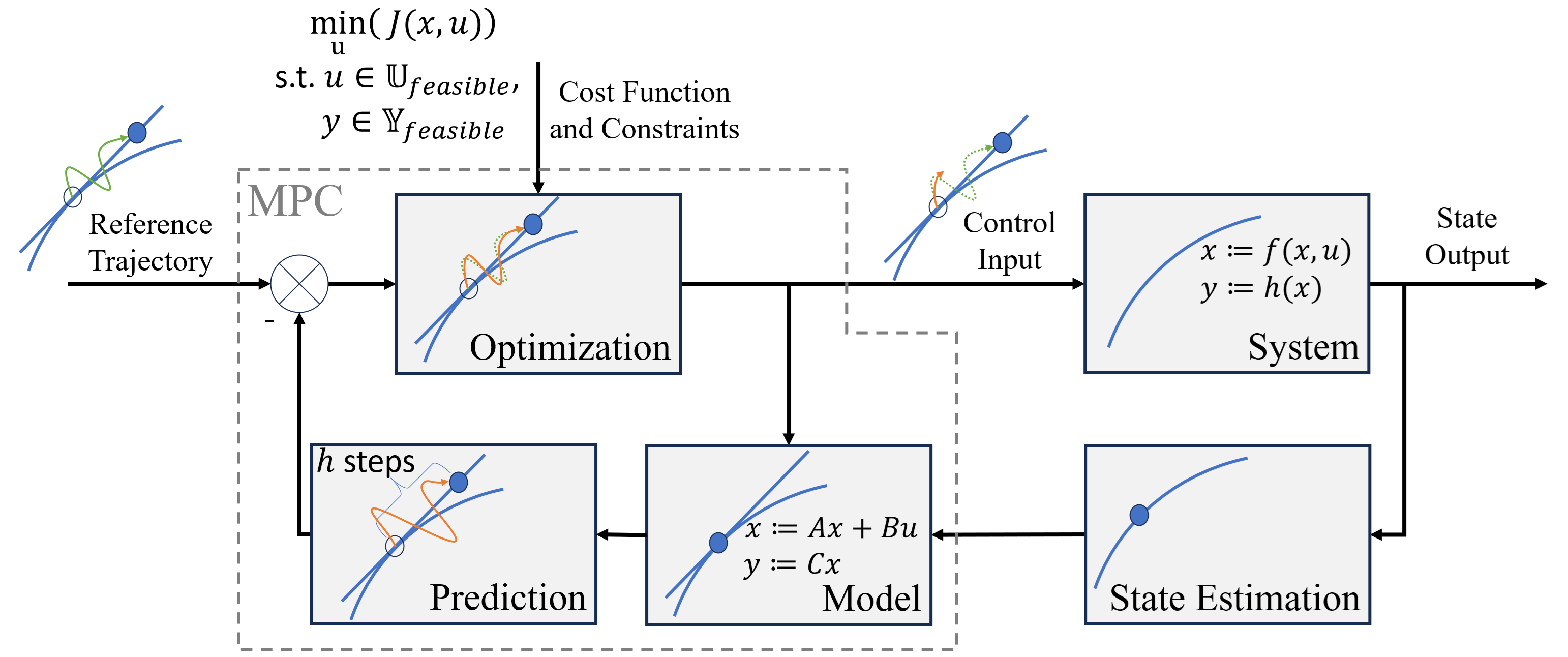}
    \caption{Block diagram of a control loop with MPC, modified from \cite{Schwenzer21}. In the MPC block, a general optimization loop produces a single control input for the current time step, though the predicted trajectory may extend several time steps into the future.}
    \label{fig:MPC_blockdiagram}
\end{figure}

MPC is a class of controllers that, using a mathematical model of a system of interest, predicts in real-time the future of that system \cite{RICHALET78}.  
One or more of these futures may be evaluated according to a cost function, or computed explicitly, and from the best future the achieving control input will be selected \cite{Bemporad00}. 
After a control input is committed to by the controller, prediction begins again, and the cycle described by Fig. \ref{fig:MPC_blockdiagram} continues. 

MPC was originally designed to offer a more system-agnostic control scenario than contemporary methods such as PID. 
These challenging control problems, although not the majority in practice, often exhibit nonlinearity, a large number of variables, time-varying dynamics, control constraints, or other special requirements or objectives related to 'optimality' \cite{RICHALET93, LEE99}. 
It is emphasized that PID handles simple systems well and is generally the appropriate choice in such cases, but MPC satisfies a need for proactive, robust control when the need arises. 

\subsection{Core Principles}
MPC can be conceptualized as these three principles: (1) System Modeling, (2) The Receding Horizon, and (3) Constrained Optimization. 
To reiterate these principles, MPC's control input is that which achieves minimal cost from constrained optimization, with a scope over a predicted receding horizon, utilizing the system's model.

System modeling is the most significant principle of MPC. 
In essence, this involves capturing the real world into a single, tangible form for prediction and optimization algorithms, which can be expressed as in Equation \ref{eq:MPC_system_formulation}. 
\begin{equation}
\begin{split}
x_{k+1} & = f(x_k, u_k) \\
y_{k} & = h(x_k)
\end{split}
\label{eq:MPC_system_formulation}
\end{equation}
where $x_k, u_k, y_k$ represent the state internals, control input, and state output at time step $k$, respectively. 
The model is the basis of MPC, in which a virtual mirror of the system is simulated to find the best-achieving inputs.

Next, the receding horizon is the extent of MPC prediction. The horizon does not need to be long-term to track a long-term trajectory, but "long enough to represent the effect of a change in the manipulated variable $u$ on the control variable $y$" \cite{Schwenzer21}.

In this case, consider a prediction horizon of $h$ time steps, where at any time point $k$, model prediction spans points $k$ to $k+h$. Likewise, MPC considers states $x_k$ to $x_{k+h}$, outputs $y_k$ to $y_{k+h}$, and a desired reference trajectory $r_k$ to $r_{k+h}$. 

Lastly, the criteria of optimality must be defined. This is practically always done with a cost function considering the weight of actions $u_k$ to $u_{k+h-1}$ and the deviation of state output from the reference. We may formulate one as in Equation \ref{eq:MPC_cost_function}.
\begin{equation}
\begin{split}
J(x_k, u) & = \sum_{i=0}^{h} \left((r_{k+i}-y_{k+i})^\intercal W_y(r_{k+i}-y_{k+i})\right) \\
          & + \sum_{i=0}^{h-1} u_{k+i}^\intercal W_u u_{k+i}
\end{split}
\label{eq:MPC_cost_function}
\end{equation}

User-defined diagonal matrices $W$ control the relative importance of trajectory tracking against minimizing $u$. 

Lastly, this objective function $J$ is minimized subject to arbitrary constraints to $x$ or $u$. These constraints may be 'hard,' or inviolable, which introduces the risk of the optimization problem being infeasible. Alternatively these constraints may be replaced by a slack variable $\xi$ and integrated into $J$.

\subsection{Challenges}
Although MPC provides a powerful framework for process control, its practical application to complex manufacturing processes faces several limitations.  
This subsection discusses three key challenges that restrict the effectiveness of MPC in such settings.

\begin{enumerate}

%Reliance on simplified pre-defined modeling that cannot fully capture the dynamic and nonlinear relationships that exist in the process.
\item \textbf{Difficulty in Modeling Complex System Behaviors.}
MPC fundamentally relies on an accurate dynamic model of the system to predict its future behavior and compute optimal control actions.  
Traditionally, such models are predominantly derived from fundamental theories or first-principles, which themselves are often established under simplifying assumptions about system conditions.  
Because it is inherently difficult to develop fully accurate theoretical models for complex manufacturing systems, these models are frequently simplified or linearized to ensure computational tractability \cite{mayne2014model, wu2025tutorial}.  
However, such simplifications cannot comprehensively capture the nonlinearities and complexities of real processes, leading to deviations from actual system dynamics, reduced prediction accuracy, and ultimately degraded control performance.  
This limitation necessitates the development of alternative modeling approaches that can better capture the dynamic and nonlinear relationships present in manufacturing processes.

% MPC requires accurate current state information, but extracting from a high-dimensional or nonlinear system is challenging.
\item \textbf{Difficulty in Estimating High-Dimensional States}
One crucial requirement of MPC is the availability of accurate estimates of the current system state.
To obtain such estimates, MPC frameworks typically rely on data from various sensors.
However, in manufacturing processes, state estimation is particularly challenging due to the high-dimensional and nonlinear nature of sensor signals and process variables \cite{gunasegaram2024machine}.  
Directly incorporating raw high-dimensional data into the MPC framework substantially increases the computational burden and exacerbates issues with measurement noise and feature redundancy, thereby degrading prediction and control performance \cite{mayne2014model, qin2003survey}. 
This challenge highlights the necessity of techniques that can extract compact and informative representations of system states from complex process data \cite{mayne2014model}.

%MPC simplifies or doesn’t consider uncertainty in its modeling.
\item \textbf{Difficulty in Handling Uncertainty.}
Conventional MPC has a limitation in handling uncertainties.  
Because the system models used in MPC are often simplified or linearized, as discussed in the first challenge above.  
Discrepancies between the model and the actual process dynamics, particularly under uncertainty, inevitably arise.  
Modeling formulations of MPC are typically deterministic, relying on these nominal models while neglecting the stochastic nature of process disturbances and sensor noise.  
Although stochastic MPC variants have been proposed to explicitly incorporate uncertainty, the propagation of uncertainty is intractable in practice \cite{mesbah2016stochastic}.  
As a result, uncertainty is frequently addressed only through simplified assumptions or conservative rule-based safety margins, which can limit control performance and adaptability under varying process conditions \cite{mayne2014model}.  
This challenge underscores the need for approaches that can more effectively represent and mitigate uncertainty within the MPC framework.

\end{enumerate}

Together, these challenges demonstrate the need for more advanced approaches that can overcome the limitations of conventional MPC in complex and uncertain manufacturing environments.

\begin{comment}

% From here - MAX's manuscripts
Recent reviews \cite{BORREGGINE19}\cite{Schwenzer21}\cite{Yu24} maintain that a major challenge MPC faces is of computation. Calculating optimal control trajectories in an on-line fashion, while made feasible with advances in modern computing, is difficult when facing high-dimensional non-linear systems and fast control intervals. This trade-off between model quality and ability to model in real-time is dominant in controller design.

Schwenzer et. al.'s review \cite{Schwenzer21} has collected ML-adjacent proposals to address these hurdles. A general application of neural nets (NNs) is to make a fast approximation of either the model or the optimal solution space for nonlinear MPC, which reduced computation time many-fold at the cost of a long one-time training session. 

Another consideration posed \cite{Yu24} is weighting factor design. When multiple control objectives are involved, the designer must tune weighting factors $W$ as appear in eq. \ref{eq:MPC_cost_function} to prioritize some factors over others. However, this is often done with trial and error, and is very time-consuming. Learning-based methods have been proposed to automate this process, though the need for intelligence in weight tuning remains a challenge to meet.
    
\end{comment}

%%%%%%%%%%%%%%%%%%%%%%%%%%%%%%%%%%%%%%%%%%%%%%%%%%%%%%%%%%%%%%%%%%%%%%%%%%%%%%%
\section{ML-driven MPC}
MPC methods have provided foundational approaches to managing manufacturing processes, but as discussed in the previous section, MPC faces significant challenges in capturing nonlinear dynamics and estimating high-dimensional states \cite{hespanha2003overcoming, fang2022process}.  
To address these limitations, ML provides complementary capabilities through data-driven modeling and latent representation learning.  
This section first outlines the core capabilities of MLs, then examines how these capabilities enhance MPC in manufacturing contexts, and finally discusses the remaining challenges of ML-driven methods.

\subsection{Capabilities of MLs}
%Explain ML's core capabilities that complement MPC limitations.
ML contributes capabilities that extend the reach of MPC beyond its traditional limitations.  
In particular, data-driven approaches help MPC better accommodate nonlinear system dynamics, while representation learning enables more tractable and robust state estimation in high-dimensional settings.

\begin{enumerate}
% (1) ML’s data-driven modeling of non-linear systems.

\item \textbf{Capturing Nonlinear and Complex Relationships through Data-Driven Learning.}
MPC often employs simplified or linearized models to maintain computational tractability, which restricts its ability to capture the nonlinear behaviors commonly observed in manufacturing systems.  
ML provides a data-driven approach by directly learning complex input–output relationships from observed data.  
ML techniques, such as neural networks (NNs), in particular, excel in discriminative modeling and prediction, using layered nonlinear activation functions to approximate intricate mappings between process inputs and outputs \cite{bishop2006pattern}.  
In the context of MPC, these mappings can correspond to process states and optimal control actions, enabling precise parameter adjustments tailored to current operating conditions without requiring a fully specified first-principles model \cite{inyang2022learn, kuang2020precise, moradimaryamnegari2025neural}.  
This capability allows ML to represent dynamics that are difficult to formulate analytically, thereby complementing MPC’s limitations in handling nonlinear and complex system behaviors.

% (2) ML learns latent representations.
\item \textbf{Latent Representation Learning.}
ML offers an effective strategy through representation learning, which provides an alternative to directly handling raw high-dimensional sensor data in MPC.  
ML can transform complex, noisy, and redundant process data into compact latent spaces that preserve the most informative features \cite{bishop2006pattern}.
These latent features primarily capture correlations within the data, and make state estimation more tractable by reducing dimensionality and computational complexity, while also improving robustness against measurement noise.  
Leveraging such representations enables more reliable predictions and control decisions, thereby addressing the challenges of high-dimensional state estimation\cite{watter2015embed}.  
More specifically, feature-based learning significantly enhances control effectiveness, as ML methods can efficiently leverage relevant features, enabling targeted control decisions without requiring full modeling of all process variables \cite{humfeld2021machine}.

\end{enumerate}
These capabilities illustrate how ML can compensate for the limitations of MPC by capturing nonlinear relationships and enabling more tractable state estimation.

\subsection{ML-driven MPC in Manufacturing}
This subsection reviews representative studies that integrate ML into MPC frameworks, with a particular focus on manufacturing applications. 
We first discuss a case study that demonstrates how ML can be embedded across different MPC components. 
Building on this example, we then synthesize insights from other works to highlight the distinct roles ML has played within MPC. 
This perspective helps us understand not only how ML has been applied so far, but also how such insights may guide the incorporation of emerging approaches into MPC in the future.

One representative case is the work by Knaak et al., which presents an ML-driven MPC framework in the context of LPBF, where ML is applied across multiple components of the MPC \cite{knaak2021improving}. 
The study addresses the challenge of improving build quality, with surface roughness serving as a key indicator. 
High dynamic range (HDR) optical images are employed as input signals, but the roughness itself is not directly available to the controller. 
To extract this feature, a convolutional NN (CNN) is used to estimate surface roughness from HDR images, thereby augmenting the state estimator in the MPC framework.

The estimated roughness features are then integrated into the predictive model. 
Instead of relying on detailed process modeling, a random forest (RF) regression model is employed as a surrogate to predict the dynamics of layer-to-layer quality. 
By training on process data, the RF provides a practical predictive model that captures the nonlinear characteristics of the LPBF process.

\begin{figure*}[htbp]
    \vspace*{10pt}
    \centering
    \includegraphics[width=\textwidth]{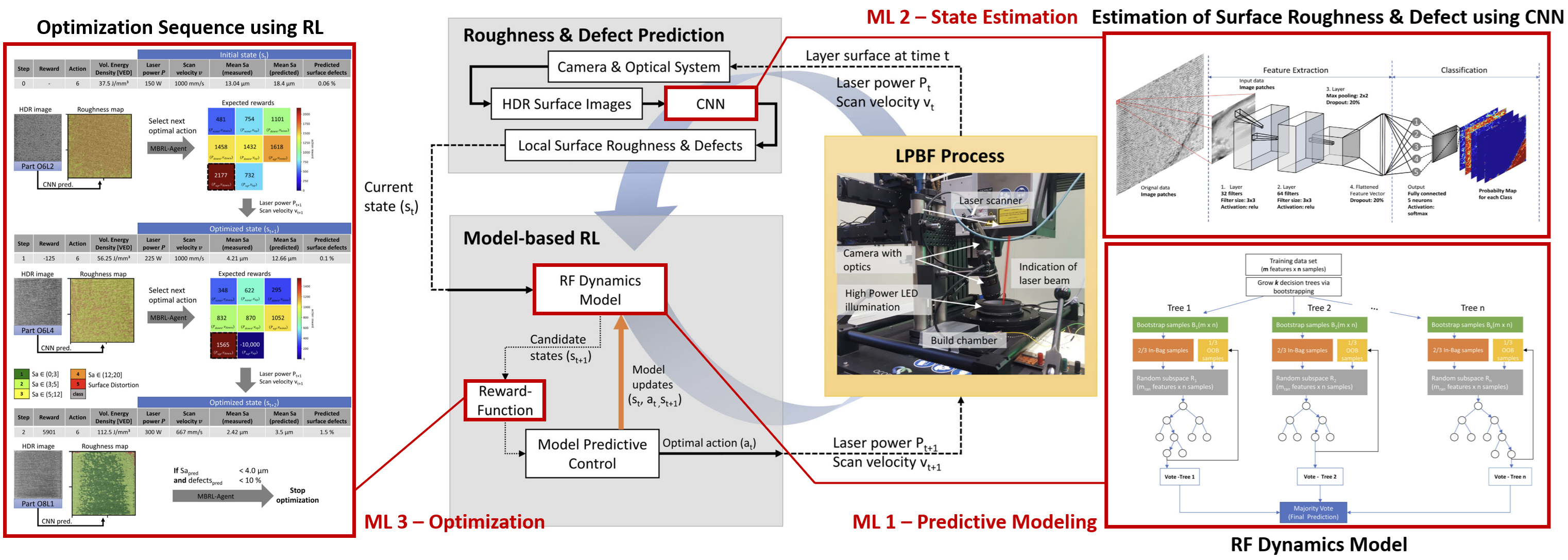}
    \caption{Reconstructed framework of the ML-driven MPC based on figures from Knaak et al.\cite{knaak2021improving}. 
    This illustration combines elements from their original figures to highlight how ML components operate within the overall MPC framework: 
    a Random Forest (RF, ML 1) serves as a surrogate predictive model for layer-to-layer quality dynamics (predictive modeling), 
    a Convolutional Neural Network (CNN, ML 2) estimates surface roughness from images (state estimation), 
    and reinforcement learning (RL, ML 3) replaces the traditional solver to optimize process parameters (optimization).}
  \label{fig:knaak_framework}
\end{figure*}

Finally, the optimization stage departs from traditional solvers by incorporating reinforcement learning (RL). Using the predictions from the surrogate model, RL policies are trained to adaptively select process parameters such as laser power and scan velocity, guided by a reward function that penalizes surface roughness and defect formation. 
The overall structure of this framework is summarized in Figure \ref{fig:knaak_framework}.
This framework illustrates how ML can simultaneously contribute to state estimation, predictive modeling, and optimization within MPC.

% ML's role in predictive modeling
Predictive modeling has been the area where ML has found the widest application among the different components of MPC. 
In reviewed studies, ML models are primarily used as surrogates to approximate complex and nonlinear process dynamics that are difficult to capture with detailed mechanistic models or computationally expensive simulations. 
Hu et al. present an approach that employs recurrent NNs (RNNs) to model nonlinear chemical processes, with updates enabling the controller to adapt across scheduled mode transitions \cite{hu2023online}. 
Similarly, physics-informed RNNs have been proposed to integrate physical prior knowledge into the surrogate model, thereby improving robustness under parameter uncertainty \cite{zheng2023physics}.
The Extreme Learning Machine (ELM), a feedforward NN-based model, serves as the predictive model in the MPC to forecast the future crystal diameter based on collected process data in Wan et al.'s study \cite{wan2021data}.
Other works have leveraged reduced-order modeling with autoencoders and regression layers to achieve tractable MPC for high-dimensional chemical systems \cite{zhao2022machine}. 
NNs have been adopted as efficient surrogates in semiconductor thin-film deposition \cite{kimaev2019nonlinear} and biopharmaceutical cell culture processes \cite{rashedi2023machine}. 
In large-scale manufacturing such as hot rolling, NN models have also been integrated with process monitoring to capture time delays and disturbances while predicting rolling dynamics \cite{xu2024novel}.
% Summary predictive modeling
Overall, these studies demonstrate that ML-based models can enhance MPC by providing accurate predictions and representations of process dynamics across various domains, utilizing a data-driven approach. 
The primary objectives have been to improve prediction accuracy under nonlinear and uncertain conditions, and to reduce the computational burden of solving dynamic models.
This widespread use of ML for predictive modeling highlights its central role in enabling MPC to handle the complexity of manufacturing processes.

% ML's role in State Estimation
Another important role of ML within MPC is state estimation, particularly when critical quality variables are not directly measurable during manufacturing. 
In semiconductor processes such as Czochralski silicon single-crystal growth, ML methods have been used to infer hidden quality indicators such as the pulling rate to thermal gradient ratio. 
Wan et al. combined stacked autoencoders with random forest regression for this task \cite{wan2021data}.
% ML State Estimation - Working as VM, soft sensor
This study illustrates how ML can be used as soft sensors, also referred to as virtual metrology.
Because MPC relies on accurate state information for prediction, ML transforms readily available process signals into estimates of critical quality metrics that reflect the current state and would otherwise require specialized measurement equipment. 
Especially in semiconductor manufacturing, where direct measurements are often expensive, slow, or impossible during operation, ML-based state estimation has gained significant traction \cite{tin2022virtual}.
% ML State Estimation - Handling high-dimensional data
In addition, ML has also been employed to handle high-dimensional sensing data by learning representations that are more informative for control. 
As shown in the study of Knaak et al., ML converts high-dimensional images into features representing surface roughness, enabling the MPC framework to monitor a current state that could not be directly obtained from raw images \cite{knaak2021improving}.
% ML State Estimation - Summary
Considering those works together, ML enables state estimation in MPC to incorporate both hidden quality indicators and features extracted from high-dimensional data.

% ML's role in Optimization
ML has also been applied at the optimization stage, where the control actions are determined.
Hosseinionari and Seethaler proposed a hybrid MPC–deep reinforcement learning (DRL) framework in the thermoforming process \cite{hosseinionari2024integration}.
In the framework, MPC is responsible for prediction and constraint handling, while DRL learns a control policy to optimize heater power and timing.
ML plays a direct role in optimization by either replacing the traditional solver with a planning mechanism, as demonstrated by Knaak et al., or by directly executing a learned policy to govern control actions, replacing the need for the MPC to run online optimization, enabling efficient process control under nonlinear and constrained conditions \cite{knaak2021improving, hosseinionari2024integration}.

In contrast, other works have shown that ML can also contribute to optimization in a more indirect manner. 
For example, Zheng and Wu introduced physics-informed RNNs that incorporate system knowledge to improve model accuracy under parameter uncertainty \cite{zheng2023physics}. 
Although their method does not directly intervene in the optimization, by embedding physical constraints into the predictive model, the MPC optimization process is effectively supported with more reliable constraint satisfaction.
Together, these studies illustrate that ML can either take an explicit role in optimization by directly learning control policies or provide implicit support by enhancing constraint handling through improved predictive modeling. Both directions expand the scope of how ML can interact with the optimization component of MPC.

% ML's Role Comprehensive Summary
In summary, the reviewed studies demonstrate that ML can contribute to MPC across different components, including predictive modeling, state estimation, and control optimization. 
In predictive modeling, ML serves primarily as a surrogate to capture nonlinear dynamics. 
In state estimation, ML enables the construction of virtual metrology tools that provide access to hidden quality variables or capture current state features from high-dimensional data. 
In optimization, ML plays both direct and indirect roles, directly through RLs that replace traditional solvers, and indirectly by enhancing constraint satisfaction via physics-informed models. 
Table \ref{tab:ml_mpc_examples} summarizes the key applications, roles, methods, and integration styles identified in this review.

%%%%% TABLE %%%%%%

\begin{sidewaystable*}[p]
  \centering
  \small
  \setlength{\tabcolsep}{5pt}
  \renewcommand{\arraystretch}{1.5}
  \caption{Integration of Machine Learning into Model Predictive Control: Representative Studies in Manufacturing}
  \label{tab:ml_mpc_examples}
  \begin{tabular}{
    >{\raggedright\arraybackslash}p{2.3cm}  % Reference
    >{\raggedright\arraybackslash}p{2.2cm}  % Application
    >{\raggedright\arraybackslash}p{2.7cm}  % ML Types
    >{\raggedright\arraybackslash}p{2.3cm}  % Role in MPC
    >{\raggedright\arraybackslash}p{4.0cm}  % Objectives (ML’s purpose)
    >{\raggedright\arraybackslash}p{2.2cm}  % Integration style
    >{\raggedright\arraybackslash}p{2.7cm}  % Control parameter(s)
    >{\raggedright\arraybackslash}p{2.7cm}  % Monitoring variables
  }
  \toprule
  \textbf{Reference} & \textbf{Application} & \textbf{ML Types} & \textbf{Role in MPC} & \textbf{Objectives (ML’s purpose)} & \textbf{Integration style} & \textbf{Control parameter(s)} & \textbf{Monitoring variables} \\
  \midrule

  Knaak et al. (2021) \cite{knaak2021improving} & LPBF & Convolutional NN (CNN) & State estimation & Generate surface roughness map from HDR images & Augmentation \newline (soft sensor) & Laser power; scan velocity & HDR coaxial images \\
  &  & Random Forest (RF) & Predictive model & Approximate layer-to-layer quality dynamics & Surrogate model & Laser power; scan velocity & Roughness; defect stats \\
  &  & Reinforcement Learning (RL) & Optimization & Learn policy to minimize roughness/defects & RL replaces numerical solver & Laser power; scan velocity & Predicted quality indicators \\

  Hu et al. (2023) \cite{hu2023online} & Chemical process & Recurrent NN (RNN) & Predictive model & Capture dynamics across modes; adapt online during transitions & Surrogate model & Reactor input flows; temperature control & Reactor state variables \\

  Zheng \& Wu (2023) \cite{zheng2023physics} & Chemical process & Physics-informed RNN & Predictive model (supports constraint satisfaction) & Physics-consistent surrogate dynamics under parameter uncertainty & Surrogate model & Feed rates; reactor temperature & Process states (x, u) \\

  Wan et al. (2021) \cite{wan2021data} & Semiconductor (crystal growth) & Stacked autoencoder; RF & State estimation & Estimate hidden quality variable (V/G) from measured process data & Augmentation \newline (soft sensor) & Pulling rate; heater power & Melt temperature; interface position; crystal diameter \\
  &  & Extreme Learning Machine (ELM) & Predictive model & Predict future output (crystal diameter) & Surrogate model & Heater power & Crystal diameter (output quantity) \\

  Kimaev~\& Ricardez-Sandoval (2019) \cite{kimaev2019nonlinear} & Semiconductor (deposition) & ANN & Predictive model & Replace computationally expensive multiscale physics model & Surrogate model & Gas flow rates; substrate temp; pressure & Film thickness; chamber process variables \\

  Zhao et al. (2022) \cite{zhao2022machine} & Chemical process & Autoencoder; RNN & Predictive model & Reduced-order modeling for fast and tractable MPC & Surrogate model & Reactor feeds; heating \& cooling & Process states (temperature; concentration) \\

  Rashedi et al. (2023) \cite{rashedi2023machine} & Biopharma cell culture & Feed-forward NN; Gaussian process (GP); linear regression & Predictive model & Approximate nonlinear cell kinetics for MPC & Surrogate model & Glucose feed rate; pH; dissolved oxygen & Cell density; metabolites; pH; DO \\

  Xu et al. (2024) \cite{xu2024novel} & Hot rolling process (steel manufacturing) & GRU–CNN & Predictive model & Capture nonlinear rolling dynamics; integrate with monitoring & Surrogate model & Roll gap; rolling speed; inter-stand tension & Rolling force; strip thickness; temperature \\

  Hosseinionari \& Seethaler (2024) \cite{hosseinionari2024integration} & Thermoforming & Deep RL & Optimization & Improve thermal tracking; energy-efficient heating & MPC-guided deep RL & Heater power; heating timing/pattern & Temperature distribution (sheet surface temperature) \\
  \bottomrule
  \end{tabular}
\end{sidewaystable*}

%%%%%%%%%%%%%%%%%%

\subsection{Challenges}
Although ML has shown promise in MPC for manufacturing processes, several challenges remain before these approaches can be broadly adopted. 
This subsection highlights three major limitations consistently observed across reviewed studies: (1) over-reliance on data, (2) limited ability to understand underlying hidden patterns because of focusing on correlation, and (3) inadequate handling of uncertainty. 
These challenges not only constrain current ML-driven MPC applications but also motivate the exploration of new paradigms in the subsequent section.

%- Over-reliance on training data, point-wise learning.
The primary challenge inherent in utilizing ML models is that their data-driven nature makes ML-driven MPC overly reliant on training data. 
Most of the reviewed studies that leverage ML for predictive modeling adopt supervised learning approaches, such as ANNs, RNNs, RF, and regression models. 
These methods learn input–output mappings, referred to as point-wise learning, from available data, which makes them effective within the training distribution but fragile when exposed to conditions not represented in the data. 
Developing an accurate and generalized ML model, therefore, necessitates a comprehensive and representative dataset. 
However, it is practically infeasible to reflect all possible scenarios in training data. 
When the ML model encounters unseen cases, it can yield inaccurate predictions, leading to degraded MPC performance. 
While some studies employ physics-informed approaches or online learning strategies, if the model fails to maintain stability characteristics derived from the initial training, MPC constraints may become recursively infeasible \cite{zheng2023physics, hu2023online}. 
As a result, the predictive capacity of these models is fundamentally tied to the availability and representativeness of training data, limiting their scalability and robustness under dynamic manufacturing conditions.

%- Limited ability to understand underlying hidden patterns, focus on correlation.
A second challenge is that ML methods integrated into MPC primarily capture statistical correlations without providing deeper insight. 
In state estimation, for example, ML-based soft sensors infer quality variables from process signals, but these estimates reflect input–output associations rather than the structural relationships that drive quality variations. 
This limitation is closely tied to the data dependence discussed earlier.
As long as models rely on statistical associations learned from training data, they remain unable to uncover deeper process relationships.
To improve the computational efficiency of MPC, reduced-order modeling approaches, often based on autoencoders, project high-dimensional measurements into compact latent variables so that MPC can be solved more quickly while preserving predictive content. 
This improves tractability; however, because the latent variables are non-invertible and lack physical meaning, important aspects of process behavior may be obscured. 
As a result, such representations make it difficult to link predictions back to process mechanisms, hindering efforts to discover deeper relationships for robust control.

%- Inadequate handling of uncertainty.
A third challenge is the limited ability of current ML-driven MPC frameworks to handle uncertainty. 
Many of the existing models are deterministic, such as ANNs and RF, providing point predictions without quantifying prediction uncertainty.
This restricts their robustness in dynamic manufacturing environments where parameter drift, unmodeled disturbances, and measurement noise are inevitable.
Uncertainty from both state prediction and estimation compounds over the prediction horizon, making MPC forecasts increasingly unreliable when uncertainty is not explicitly modeled.
Although some probabilistic models like Gaussian Processes (GPs) can offer uncertainty estimates through their probabilistic structure, and physics-informed approaches embed prior knowledge to mitigate these issues, their effect remains indirect and often insufficient to guarantee constraint satisfaction. 
Similarly, reinforcement learning–based optimizers adaptively learn control policies, but their reliance on data-driven reward functions makes them vulnerable to variability and constraint violations under unseen conditions. 
Overall, the lack of uncertainty representation and propagation in ML components undermines the reliability of MPC when deployed in complex and evolving manufacturing processes.

% Summary of ML-driven MPC challenges
Together, these challenges illustrate the limitations of current ML-driven MPC frameworks. 
Their reliance on training data restricts generalization, their focus on correlations obscures deeper process relationships, and their limited treatment of uncertainty undermines reliability under dynamic operating conditions. 
Addressing these issues requires approaches capable of learning richer data distributions, capturing hidden structures, and naturally incorporating uncertainty. These motivations set the stage for the discussion of generative ML-driven MPC in the following section.

%%%%%%%%%%%%%%%%%%%%%%%%%%%%%%%%%%%%%%%%%%%%%%%%%%%%%%%%%%%%%%%%%%%%%%%%%%%%%%%
\section{Generative ML-driven MPC}
This section introduces the core capabilities of generative ML and reviews key methods, including GANs, normalizing flows, VAEs, diffusion models, and LLMs. We then examine how these methods integrate with MPC, highlighting their respective roles, and finally synthesize how generative ML as a whole enhances predictive modeling, state estimation, and optimization within the MPC framework.

\subsection{Capabilities of Generative ML}
\label{subsection_capa_GenML}

Generative ML has emerged as a powerful tool for solving complex problems across many domains \cite{goodfellow2016deep}. 
Generative ML refers to methods that aim to learn and model the underlying data distributions from observed data. 
They enable the generation of new samples and the prediction of unseen states that reflect the learned patterns \cite{bishop2006pattern, goodfellow2016nips, bond2021deep}. 
Beyond generating samples, generative ML also uncovers hidden features and captures probabilistic dependencies within the data \cite{bishop2006pattern}. 
These aspects manifest as three main capabilities, which provide a concrete basis for understanding the role of generative ML in MPC.  

\begin{enumerate}
\item \textbf{Learning Data Distributions.}
A distinctive capability of generative ML is learning data distributions, which differentiates it from conventional ML \cite{goodfellow2016nips}. 
Conventional ML methods, typically discriminative approaches, learn direct mappings from inputs to outputs by maximizing the conditional likelihood $p(y \mid x)$, as shown in Equation~\ref{eq:disc_likelihood} \cite{murphy2012machine, goodfellow2016deep}.  

\begin{equation}
\theta^* = \arg\max_\theta \sum_{i=1}^N \log p(y_i \mid x_i;\theta)
\label{eq:disc_likelihood}
\end{equation}
where $\theta^*$ represents the optimal model parameters obtained via maximum likelihood estimation of the conditional likelihood, $i$ indexes the data samples, and $N$ denotes the total number of data.  
ML’s discriminative learning thus primarily captures correlation-based relationships, without modeling the underlying data distribution.  

By contrast, generative ML learns a model distribution $p_{\text{model}}(x;\theta)$ that approximates the real data distribution $p_{\text{data}}(x)$, as represented in Equation~\ref{eq:gen_likelihood} \cite{murphy2012machine, goodfellow2016deep}.  

\begin{equation}
\theta^* = \arg\max_\theta \sum_{i=1}^N \log p_{\text{model}}(x_{i};\theta)
\label{eq:gen_likelihood}
\end{equation}

This distributional perspective is particularly valuable in MPC. 
Whereas conventional ML-driven MPC often relies on deterministic mappings (e.g., $\hat{x}_{t+1} = f_\theta(x_t,u_t)$), generative models allow probabilistic solution spaces (e.g., $x_{t+1} \sim p_\theta(x \mid x_t,u_t)$), providing a richer representation of system dynamics. 
Additionally, this approach enables models to learn effectively under limited data availability, thereby enhancing predictive models of system dynamics for MPC.

\item \textbf{Capturing Hidden Patterns and Relationships.}
Based on the distribution learning capability, generative ML facilitates the uncovering of hidden patterns and relationships embedded within those distributions \cite{goodfellow2016deep}.  
While conventional ML uses latent features as internal representations to support input-output mapping, generative ML introduces latent variables that represent probabilistic hidden factors explaining how data are generated \cite{bishop2006pattern, goodfellow2016deep}.  
As shown in Equation \ref{eq:latent}, generative ML assumes data arise from underlying latent variables \cite{bishop2006pattern, kingma2013auto}.

\begin{equation}
p_\theta(x) = \int p_\theta(x \mid z)\,p(z)\,dz
\label{eq:latent}
\end{equation}
The latent variable $z$ captures hidden relationships, and the conditional distribution $p_\theta(x \mid z)$ describes how these hidden factors manifest in the observed data.  
By marginalizing over all latent variables, the model recovers the data distribution $p_\theta(x)$, thereby uncovering hidden relationships that govern data generation.  

Additionally, generative ML naturally handles multi-modal distributions \cite{goodfellow2016nips}.  
This ability enables the identification of distinct modes within a dataset, which helps reveal different patterns of system behavior that may not be captured by traditional correlation-based analysis~\cite{goodfellow2016deep}.  
Such multi-mode modeling also allows the same input conditions to correspond to multiple plausible outcomes, thereby enriching the solution space that generative ML can represent.  

For MPC, this capability means that generative models not only represent diverse possible solutions but also reflect the dependencies underlying the solution space, offering a more faithful basis for predictive control than purely discriminative mappings.

\item \textbf{Handling Uncertainties.}
Generative ML also provides a natural way to handle uncertainty through distribution learning.  
It represents state transitions as probability distributions, which enables generative ML to capture expected outcomes as well as the variability around them.  
Such probabilistic representations reflect the inherent uncertainty of the system and enable the consideration of diverse scenarios that deterministic predictions cannot capture.  

For MPC, such uncertainty modeling is essential for robust and risk-sensitive decision making\cite{mesbah2016stochastic}.  
Generative ML enables the design of predictive models that incorporate variability directly into the optimization process and support robust MPC formulations that are not possible with purely deterministic mappings.  
In particular, generative ML can support efficient uncertainty propagation by characterizing full probability distributions rather than relying on conservative approximations\cite{mesbah2016stochastic}.
\end{enumerate}

\subsection{Key Generative ML Methods}
\label{subsection:Key Gen ML Methods}
As discussed in Section \ref{subsection_capa_GenML}, one of the main capabilities is that generative ML learns the data distribution of the training dataset.  
Based on this learning perspective, Goodfellow proposed a taxonomy of generative ML methods, which distinguishes between explicit and implicit approaches, depending on how the model understands the distribution of data and represents the likelihood function \cite{goodfellow2016deep}.  
Explicit density methods can express the density function of input data as a likelihood function $p_{\text{model}}(x; \theta)$ through a straightforward calculation.
Within explicit density methods, there are two approaches to modeling the density of data that capture underlying complex relationships.  
One is a tractable approach that defines a computationally tractable density function.  
The other approach allows approximations of the likelihood when the data has high dimensionality or nonlinearity, which are otherwise intractable to compute exactly.  
These approximate density methods can be distinguished by approximating via the variational or the Markov chain approach.  
Variational approximation induces the distribution by introducing a simpler variational distribution; on the other hand, the Markov chain approach approximates a distribution by drawing samples through a Markov chain.
Unlike the aforementioned explicit methods, implicit density methods do not explicitly represent a probability distribution.

\begin{figure}[htbp]
    \vspace*{10pt}
    \centering
    \includegraphics[width=\columnwidth]{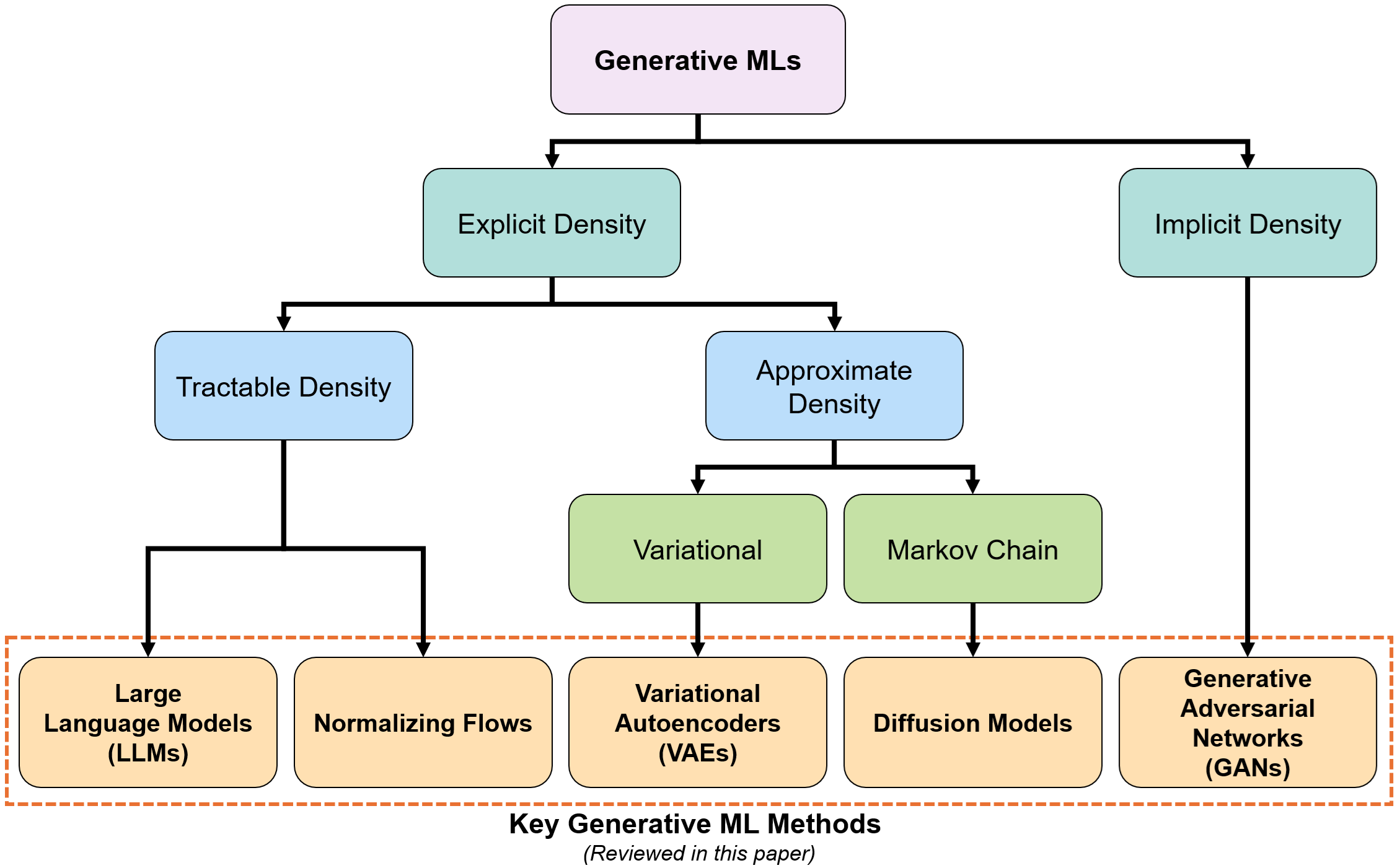}
    \caption{Taxonomy of generative ML methods adapted from Goodfellow et al. \cite{goodfellow2016deep}. 
    The taxonomy distinguishes between explicit and implicit density methods based on whether the data distribution is modeled through an explicit likelihood function. 
    Explicit methods include tractable approaches and approximate approaches, which are further divided into variational and Markov chain-based methods. 
    Implicit methods, by contrast, do not define a likelihood function explicitly but instead learn to generate samples directly. 
    The bottom layer highlights the key generative ML methods reviewed in this paper, indicating where each method is positioned within the taxonomy.}
    \label{fig:taxonomy}
\end{figure}

In this review, we focus on key generative ML methods following the taxonomy introduced by Goodfellow et al.\cite{goodfellow2016deep}.
We begin our review with Generative Adversarial Networks (GANs), the most prominent representative methods of implicit density models, which generate data without explicitly defining a likelihood function \cite{goodfellow2014generative}.
We then examine explicit density methods, starting with tractable approaches, the Normalizing Flows (NFs).
For explicit approximate density methods, we consider two major approaches.  
The variational method is represented by Variational Autoencoders (VAEs).  
The Markov chain method has included energy-based models such as Boltzmann Machines.  
In this review, however, we focus on Diffusion models, which can be interpreted as Markov chain-based approaches and have recently gained prominence as effective generative methods across domains \cite{yang2023diffusion}.
Finally, we include Large Language Models (LLMs).  
While LLMs belong to the class of autoregressive models under explicit tractable density methods, we examine them separately due to their significance as large-scale sequence models, which illustrates the potential of autoregressive generative methods for complex sequential decision-making tasks \cite{yang2023foundation, wang2024large}.
The overall taxonomy and the key methods reviewed in this paper are summarized in Figure \ref{fig:taxonomy}.

%%%%%%%%%%%%%%%%%%%%%%%%%%%%%%%%%%%
\subsubsection{Generative Adversarial Networks (GANs)}
GANs are implicit generative ML methods designed to generate new samples that closely resemble real data \cite{goodfellow2014generative}. 
GANs consist of two NNs, a generator and a discriminator, trained in a competitive framework.
The generator learns to produce synthetic data that mimics the training distribution, while the discriminator attempts to distinguish between real and generated samples.
During training, the generator aims to improve its output to fool the discriminator, while the discriminator tries to classify real data versus fake data correctly.
This adversarial process does not require an explicit likelihood function, and is formalized in Equation \ref{Eq:gan_objective}.
%%%%%%%%%%%%%%%%%%%%%%%
\begin{multline}
\min_G \max_D V(D, G) = \mathbb{E}_{\mathbf{x} \sim p_{\text{data}}(\mathbf{x})}\left[\log D(\mathbf{x})\right] \\
+ \mathbb{E}_{\mathbf{z} \sim p_{\mathbf{z}}(\mathbf{z})}\left[\log(1 - D(G(\mathbf{z})))\right]
\label{Eq:gan_objective}
\end{multline}
\noindent where $G$ represents the generator, $D$ is the discriminator , $p_{\text{data}}(\mathbf{x})$ is the real data distribution, and $p_{\mathbf{z}}(\mathbf{z})$ is the prior noise distribution, typically Gaussian.
%%%%%%%%%%%%%%%%%%%%%%%

In MPC contexts, GANs offer two distinctive properties:

\begin{enumerate}
    \item \textbf{Adversarial distribution learning.}
    GANs employ an adversarial training mechanism where the discriminator acts as an adaptive loss function, driving the generator to capture complex relationships between process variables without requiring an explicit likelihood~\cite{goodfellow2014generative}. 
    This implicit approach is especially practical for modeling high-dimensional process data where explicit probabilistic modeling is infeasible. 
    By learning such distributions, GANs generate diverse synthetic trajectories that support the predictive models in MPC.

    \item \textbf{Fault and anomaly scenario generation.}
    GANs' adversarial mechanism also facilitates the generation of rare or anomalous process trajectories, as the discriminator forces the generator to adapt even to subtle deviations observed in real data~\cite{sabuhi2021applications, kumbhar2023deepinspect}. 
    Such synthetic fault scenarios allow MPC controllers to be tested under abnormal conditions, improving their robustness and reliability in manufacturing environments~\cite{wang2025mcgan}.
\end{enumerate}

%%%%%%%%%%%%%%%%%%%%%%%%%%%%
\subsubsection{Normalizing Flows (NFs)}
NFs are generative methods that represent the data distribution as an explicit and tractable density.  
Let $z$ be a latent variable, and let $p_\theta(z)$ be a tractable density, typically chosen as a simple distribution such as a Gaussian (e.g., $z \sim \mathcal{N}(0, I)$).  
The generation process for data $x$ can be defined with a function $g$ as in Equation~\ref{eq:nf_generation}:  

\begin{equation}
z \sim p_\theta(z), \quad x = g_\theta(z)
\label{eq:nf_generation}
\end{equation}
where the function $g$ is invertible, in other words, bijective. Therefore, the latent variable $z$ can also be obtained from data $x$ by the inverse function $f$, as in Equation \ref{eq:nf_inverse}:  

\begin{equation}
z = f_\theta(x) = g_\theta^{-1}(x)
\label{eq:nf_inverse}
\end{equation}

The invertibility property is crucial, as it preserves all information during transformations, allowing for exact likelihood evaluation and bidirectional conversion between complex data distributions and simple latent representations.  

The function $f$ is typically defined as a sequence of transformations as in Equation \ref{eq:nf_sequence}, which can equivalently be expressed in terms of intermediate variables as in Equation \ref{eq:nf_bidirectional}.
\begin{equation}
f = f_K \circ f_{K-1} \circ \cdots \circ f_1
\label{eq:nf_sequence}
\end{equation}  
\begin{equation} x \xleftrightarrow{f_1} h_1 \xleftrightarrow{f_2} h_2 \cdots \xleftrightarrow{f_K} z \label{eq:nf_bidirectional} \end{equation}
where each $h_i$ denotes an intermediate latent representation.  
Such a sequence of invertible transformations is called an NF.  
Through the change-of-variables rule, the density of $x$ can be expressed as Equation \ref{eq:nf_changevar}:
\begin{equation}
p(x)=p(z)\,\left|\det \frac{\partial f_\theta(x)}{\partial x}\right|
\label{eq:nf_changevar}
\end{equation}

This formulation enables learning complex data distributions $p(x)$ from a simple base distribution $p(z)$~\cite{kingma2018glow, kobyzev2020normalizing, bond2021deep}.

In MPC contexts, NFs provide distinctive advantages:

\begin{enumerate}
\item \textbf{Exact likelihood for probabilistic state estimation.}
Unlike implicit generative models, NFs enable exact likelihood evaluation through the change-of-variables formulation \cite{rezende2015variational, kobyzev2020normalizing, papamakarios2021normalizing}.
It means that the probability of a given process state can be computed exactly rather than approximated. 
As a result, the uncertainty associated with each state estimate can be quantified more precisely, which is crucial when MPC must make predictions and optimize control inputs under uncertain manufacturing conditions.

\item \textbf{Invertible mapping for bidirectional process modeling.}
NFs are composed of invertible transformations, which makes it possible to move back and forth between latent representations and observed process data \cite{kingma2018glow, bond2021deep}. 
In MPC, the inverse mapping from latent space to data corresponds to prediction, where process parameters or latent variables are translated into possible future states and trajectories. 
The forward mapping from data to latent space corresponds to inference, where observed states are compressed into latent representations that capture underlying process parameters or hidden factors. 
This bidirectional capability enhances MPC by supporting both predictive control and diagnostic analysis, especially in manufacturing.
\end{enumerate}

%%%%%%%%%%%%%%%%%%%%%%%%%%%%
\subsubsection{Variational Autoencoders (VAEs)}
VAEs represent an early and influential advance in generative models that leverage variational inference and combine it with deep NNs~\cite{kingma2013auto}, and are categorized as explicit approximate density models.
The ``variational'' aspect refers to approximating the intractable posterior $p(z|x)$ with a tractable distribution, typically a Gaussian, so that inference becomes feasible.

VAEs parameterize the latent distribution conditioned on the input data with mean and variance, and then reconstruct the data through a decoding process.
%%%%%%%%%%%%%%%%%%%%%%%%%%%%%%
The encoder models input data as a probability distribution in latent space, and employs the reparameterization trick, as shown in Equations~\ref{eq:vae_encoder} and \ref{eq:vae_reparam}.
\begin{equation}
q_\phi(z|x) = \mathcal{N}(z; \mu_\phi(x), \sigma_\phi^2(x) I)
\label{eq:vae_encoder}
\end{equation}
\begin{equation}
z = \mu_\phi(x) + \sigma_\phi(x) \cdot \epsilon, \quad \epsilon \sim \mathcal{N}(0, I)
\label{eq:vae_reparam}
\end{equation}
\noindent where $x$ is the input data, $z$ is the latent variable, $\mu_\phi(x)$ and $\sigma_\phi^2(x)$ are the mean and variance outputs from the encoder network, and $\epsilon$ is random noise sampled from a standard normal distribution.
This approach enables optimization of the evidence lower bound, as shown in Equation~\ref{eq:vae_elbo}.
\begin{equation}
\mathcal{L}(\theta, \phi) = \mathbb{E}_{q_\phi(z|x)} [\log p_\theta(x|z)] 
- D_{\text{KL}}(q_\phi(z|x) \Vert p(z))
\label{eq:vae_elbo}
\end{equation}
\noindent where $\theta$ and $\phi$ are the decoder and encoder parameters, $\mathbb{E}$ denotes expectation, $p_\theta(x|z)$ is the decoder likelihood, and $q_\phi(z|x)$ is the encoder distribution. The first term maximizes reconstruction quality, while the second term with prior $p(z)$ and Kullback--Leibler divergence $D_{KL}$ serves as regularization.
%%%%%%%%%%%%%%%%%%%%%%%%%%%%%%
This probabilistic foundation enables VAEs to capture uncertainty inherently in their architecture, differentiating them from deterministic autoencoders~\cite{doersch2016tutorial}.

VAEs offer several properties that make them particularly valuable in MPC: 
\begin{enumerate}
    \item \textbf{Latent space representation for control.}
    Their latent space representation provides a useful representation for process control. 
    This latent space captures physically meaningful relationships of process dynamics, allowing a reduced-dimensional space that enables efficient control~\cite{watter2015embed}. 
    The reduction maintains critical relationships between process variables while eliminating redundant information, leading to more tractable optimization problems in MPC frameworks. 
    Notably, the learned representations often align with physically meaningful process parameters, enhancing interpretability.
    \item \textbf{Probabilistic uncertainty quantification.}
    The probabilistic framework of VAEs inherently quantifies uncertainty in the latent representation of input states~\cite{kingma2013auto}.
    This capability is crucial for robust MPC, as it enables the controller to identify regions with a higher risk of anomalous states and supports more cautious adjustments to actions. 
    At the prediction level, this property provides uncertainty estimates of model outputs, which can be leveraged to support safer decision-making in uncertain manufacturing environments~\cite{doersch2016tutorial, hewing2019cautious}.
\end{enumerate}

%%%%%%%%%%%%%%%%%%%%%%%%%%%%%%%%%%%
\subsubsection{Diffusion Models}
Diffusion models are generative models that belong to the class of explicit approximate density methods, specifically those based on Markov chains. 
They operate based on the principle of gradually denoising data through a learned reverse diffusion process \cite{ho2020denoising,nichol2021improved}.
The diffusion framework consists of two key steps: the forward process and the reverse process.

% Forward
In the forward process, the model gradually adds Gaussian noise to the original data across the steps, which means diffusing data, as shown in Equation~\ref{Eq:diff_forward}.
\begin{equation}
q(\mathbf{x}_t | \mathbf{x}_{t-1}) = \mathcal{N}(\mathbf{x}_t; \sqrt{1 - \beta_t} \mathbf{x}_{t-1}, \beta_t \mathbf{I}), \quad t \in [1,T]
\label{Eq:diff_forward}
\end{equation}
\noindent where $\mathbf{x}_0$ represents the original data, $\mathbf{x}_t$ is the noised data at timestep $t$, and $\beta_t$ is the noise schedule controlling the diffusion rate.

% Reverse
Then, the model learns the data distribution through a reverse process that reconstructs the original data using iterative denoising step by step, as defined in Equations~\ref{Eq:diff_reverse} and \ref{Eq:diff_learning}.
\begin{equation}
p_\theta(\mathbf{x}_{t-1} | \mathbf{x}_t) = \mathcal{N}(\mathbf{x}_{t-1}; \boldsymbol{\mu}_\theta(\mathbf{x}_t, t), \boldsymbol{\Sigma}_\theta(\mathbf{x}_t, t))
\label{Eq:diff_reverse}
\end{equation}
% Learning 
\begin{equation}
\mathcal{L}_{\text{simple}} = \mathbb{E}_{\mathbf{x}_0, \boldsymbol{\epsilon}, t} \left[ \|\boldsymbol{\epsilon} - \boldsymbol{\epsilon}_\theta(\mathbf{x}_t, t) \|^2 \right]
\label{Eq:diff_learning}
\end{equation}
\noindent where $p_\theta$ represents the learned reverse process with parameters $\theta$, $\boldsymbol{\mu}_\theta$ and $\boldsymbol{\Sigma}_\theta$ are the predicted mean and covariance, and $\boldsymbol{\epsilon}_\theta$ is a NN that predicts the noise component added during the forward process.

This iterative refinement allows diffusion models to generate results that are not only high in fidelity but also adaptable to process-specific constraints, making them particularly effective for tasks requiring gradual state transitions.
This approach is fundamentally different from both VAEs and GANs, as it directly models the gradient of the data distribution through score matching \cite{song2020score}.

The unique iterative nature of diffusion models offers several advantageous properties from the perspective of process control in MPC:  

\begin{enumerate}
    \item \textbf{Iterative trajectory refinement.}  
    The iterative denoising process enables process optimization by generating highly realistic process trajectories \cite{janner2022planning, giannone2023aligning, elhambakhshamdiffusion}.  
    The gradual refinement across multiple steps allows for precise control over the generation process, which differs from single-step generation approaches in other models.  
    This capability is especially valuable in planning smooth transitions between process states while adhering to physical constraints.

    \item \textbf{Probabilistic uncertainty propagation.}  
    Diffusion models provide robust probabilistic state estimation through their probabilistic formulation of the denoising reverse process \cite{tang2024adadiff}.
    Unlike VAEs, where uncertainty is mainly captured in the latent representation, diffusion models propagate uncertainty step by step across the entire denoising trajectory \cite{tang2024adadiff, janner2022planning}.
    This sequential propagation aligns naturally with the multi-step prediction horizon in MPC, enabling uncertainty to be explicitly modeled and carried forward over time \cite{mesbah2016stochastic}.

    \item \textbf{Constraint-guided risk-aware control.}  
    The denoising steps can be guided by incorporating control objectives or physical constraints, effectively combining the generative process with guidance methods so that the probabilistic estimates remain aligned with desired outcomes \cite{dhariwal2021diffusion,song2020score, bar2023multidiffusion, giannone2023aligning}.  
    This property enables risk-aware control strategies that account for varying levels of uncertainty at different stages of process evolution.  
    In addition, diffusion models can generate multiple plausible trajectories from the same initial conditions, supporting scenario analysis for control and improving reliability and adaptability in dynamic systems.
\end{enumerate}

%%%%%%%%%%%%%%%%%%%%%%%%%%%%%%%%%%%
\subsubsection{Large Language Models (LLMs)}
Large Language Models (LLMs) are examples of autoregressive models under explicit tractable density methods, which predict the next output by conditioning on the sequence of inputs. 
The autoregressive approach is based on the chain rule of probability, factorizing any joint distribution over a sequence $x = (x_1, \dots, x_n)$ in an autoregressive manner, as in Equation~\ref{eq:llm_autoreg}~\cite{bond2021deep, yang2023foundation}.

\begin{equation}
p(x) = p(x_1, \dots, x_n) = \prod_{i=1}^n p(x_i \mid x_1, \dots, x_{i-1})
\label{eq:llm_autoreg}
\end{equation}

LLMs are autoregressively trained with immense amounts of sequential data, which has led to significant performance improvements~\cite{minaee2024large, zhao2023survey}. 
The introduction of the attention mechanism in the transformer architecture has further enabled LLMs to model complex sequential dependencies at scale. 
The self-attention mechanism excels at processing sequential data, and when combined with autoregressive learning, it creates synergistic effects that allow LLMs to capture long-range dependencies effectively \cite{minaee2024large, vaswani2017attention}.

The attention mechanism enables the direct modeling of dependencies regardless of their sequential distance, as shown in Equation~\ref{Eq:Attention} \cite{vaswani2017attention}.

\begin{equation}\label{Eq:Attention}
    \text{Attention}(Q, K, V) = \text{softmax}\left(\frac{QK^T}{\sqrt{d_k}}\right)V
\end{equation}

\noindent where $Q$ represents queries as the current process state, $K$ represents keys as reference points in the data history, $V$ represents values as data associated with those reference points, and $d_k$ is the dimension of the keys. 
The resulting attention output is a weighted sum of values, where the weights reflect the relevance of each reference point to the current data point, which can be interpreted as a process state in the MPC perspective.

In MPC contexts, LLMs offer a distinctive property:  
\begin{itemize}
    \item \textbf{Sequential decision-making with long-horizon dependencies.}
    Through their attention-based autoregressive structure, LLMs excel at modeling temporal dependencies across extended sequences. 
    This capability aligns with MPC's predictive planning paradigm, where future states are anticipated over planning horizons and used to optimize control actions~\cite{yang2023foundation, wang2024large}.
\end{itemize}

\begin{comment}
\begin{enumerate}
    \item \textbf{Sequential decision-making with long-horizon dependencies.}
    Through their attention-based autoregressive structure, LLMs excel at modeling temporal dependencies across extended sequences. 
    This capability aligns with MPC's predictive planning paradigm, where future states are anticipated over planning horizons and used to optimize control actions~\cite{yang2023foundation, wang2024large}.
    %\item
\end{enumerate}
\end{comment}

%%%%%%%%%%%%%%%%%%%%%%%%%%%%%%%%%%%
\subsection{Integration Methodologies Generative ML with MPC}
This subsection examines how generative ML integrates with MPC. Generative ML-driven MPC remains an emerging area, and applications specifically targeting manufacturing are still limited. 
To address this, we review control-oriented studies where generative ML has been applied to MPC, focusing on how these models are combined, the roles they play, and the functions they enable. 
Building on the key generative methods outlined in Section \ref{subsection:Key Gen ML Methods}, we discuss how each model family has been integrated into different MPC components and the contributions they provide.
Finally, we synthesize these findings to highlight common integration patterns and reflect on how they can inform future applications of generative ML-driven MPC in manufacturing processes.

\subsubsection{Roles of GANs in MPC}
%(GANs for Predictive Modeling in MPC)
GANs enhance MPC prediction across diverse scenarios.
GANs have been integrated into MPC primarily to enhance predictive modeling.
Their ability to generate realistic sequences enables high-fidelity forecasting of complex system dynamics.
Li et al. used conditional GANs with GRUs for welding systems \cite{li2025generative}.
Their framework generates future weld pool images based on torch speed adjustments, creating a human-centered MPC system where operators visualize consequences before implementation.
The framework, enhanced with GRUs for temporal modeling, captures relationships between speed variations and weld pool morphology while preserving human judgment in the loop.
The system effectively reduces operator skill requirements by providing future-state forecasts that support decision-making.
Bae et al. introduced Social GANs (SGANs), which provide interactive trajectory forecasts of surrounding vehicles embedded into MPC collision-avoidance constraints to ensure safety in dense traffic \cite{bae2022lane}.
Yang et al. proposed an adversarial dynamics modeling approach for mobile robots, where the generator learns the system dynamics while the discriminator enforces temporal consistency and realism, effectively serving as an implicit constraint within MPC optimization \cite{yang2022sms}.
In short, GANs excel at generating realistic scenarios, such as for visual feedback, multi-agent interactions, and unstructured environments.
This enhances MPC’s predictive accuracy and enables constraint-aware decision-making in different environments.

%(GANs for Optimization in MPC)
GANs have also been leveraged to improve the optimization process in the MPC framework.
GANs enable MPC optimization through two distinct pathways.
One approach is direct policy optimization, where MPC itself becomes the generator.
Burnwal et al. proposed an adversarial training MPC approach that trains the control laws, where the control sequence computed by the MPC optimization acts as the generator and is guided by a discriminator that measures divergence from expert trajectories.
This formulation allows the MPC control law to imitate expert behaviors even under mismatched dynamics, effectively enabling the cost function to adapt automatically through imitation rather than manual design \cite{burnwal2023gan}.
Another approach is indirect reward learning, where MPC expertise guides adversarial reward function learning for RL policies.
Gupta et al. presented an adversarial inverse RL (AIRL) approach, where MPC-generated expert trajectories are used to train a GAN discriminator that serves as a reward function estimator \cite{gupta2025adversarial}.
The learned reward then guides policy optimization in RL, addressing the common challenge of manual reward specification.

Together, these studies demonstrate how GANs facilitate policy imitation, reward shaping, and distribution matching for optimization, which extends MPC from a deterministic optimizer to a more flexible and learning-driven decision-making framework.

\subsubsection{Roles of NFs in MPC}

%(Normalizing Flows for Predictive Modeling in MPC)
NFs have been applied to enhance MPC prediction by explicitly learning stochastic system dynamics.
Cramer et al. proposed a conditional NF framework that models the probability distribution of state increments given the current state and control inputs \cite{cramer2024least}.
This probabilistic representation enables scenario generation for stochastic MPC, allowing for expected least-squares optimization and marginal log-likelihood objectives, while formulating chance constraints.
Such integration highlights the ability of NFs to capture nonlinear, non-Gaussian, and state-dependent fluctuations, providing rigorous uncertainty quantification within MPC frameworks.

%(Normalizing Flows for Optimization in MPC)
NFs enhance MPC optimization through three distinct mechanisms: inverse modeling, sampling distribution learning, and latent space optimization.
Zhang and Mikelsons introduced IMFlow, a conditional NF that performs inverse modeling by mapping observations to input sequences, enabling probabilistic control input estimation and uncertainty quantification for MPC \cite{zhang2024imflow}.
For sampling-based MPC, Power and Berenson developed FlowMPPI and FlowiCEM, where a context-conditioned NF replaces the base sampling distribution of sampling-based MPC methods, MPPI and iCEM, allowing efficient trajectory sampling across environments and out-of-distribution generalization \cite{power2024learning}.
Rabenstein et al. employed NFs to enhance MPPI sampling for autonomous driving, pre-training on trajectories with heuristic rules to achieve more efficient exploration than Gaussian sampling \cite{rabenstein2024sampling}.
Sacks and Boots further advanced this line by mapping MPC optimization into the NF’s latent space, where warm-starting and online updates are more efficient, while the NF ensures that mapped control inputs automatically satisfy box constraints and feasibility \cite{sacks2023learning}.

In summary, most existing studies employ NFs primarily to improve MPC optimization rather than predictive modeling.
Collectively, they demonstrate complementary benefits: inverse modeling enables input estimation and uncertainty quantification in black-box settings, sampling distribution learning enhances efficiency and generalization of trajectory exploration, and latent-space optimization accelerates online updates while ensuring constraint satisfaction.

\subsubsection{Roles of VAEs in MPC}
%(VAEs for State Estimation in MPC)
VAEs enhance state estimation in MPC by capturing latent representations from non-measurable variables and high-dimensional observations.
Baumeister et al. applied a VAE within a deep learning–MPC architecture for mode-locked fiber lasers.
The VAE inferred latent representations of birefringence variables, which are not directly measurable, from observable laser states whenever prediction errors exceeded a threshold \cite{baumeister2018deep}.
This enabled robust control despite stochastic drifting of the underlying process.
Kwon et al. employed a Kalman VAE (KVAE) for robotic control, where the Kalman filter was embedded in the latent state space to infer compact representations from high-dimensional raw images, providing efficient initialization for MPC dynamics \cite{kwon2020combining}.
These works highlight the role of VAEs as state estimators that allow MPC to operate effectively even when key system variables are hidden or difficult to measure.

%(VAEs for Predictive Modeling in MPC)
VAEs have also been utilized to construct predictive models that approximate the dynamics of unknown or nonlinear systems.
Yao et al. proposed ControlVAE, which learns a latent dynamics model capturing unknown dynamics for physics-based character animation, enabling diverse and realistic motion control through sampling-based MPC \cite{yao2022controlvae}.
Kwon et al. utilized KVAE for a latent dynamics model, learning deterministic, time-invariant models through latent encoding in low-dimensional spaces \cite{kwon2020combining}. 
This approach enables efficient long-term prediction without autoregression by operating on augmented state representations derived from the VAE's latent space.
Estiri and Mirinejad introduced a VAE-based recurrent nonlinear state-space model (RNSSM) for clinical fluid resuscitation, where the VAE captured nonlinear relationships between physiological states and outputs, and incorporated variational Gaussian inference to handle noise and uncertainty \cite{estiri2024variational}.
These studies demonstrate how VAEs provide compact yet expressive predictive models, capable of representing stochastic and nonlinear behaviors while improving the robustness of MPC forecasts.

%(VAEs for Optimization in MPC)
VAEs have been leveraged to improve the optimization process within MPC.
In ControlVAE, high-level motion patterns were encoded as latent variables, and this latent skill representation served as a generative control policy. 
By sampling from this latent distribution rather than raw control inputs, MPC could efficiently compose and explore diverse behaviors \cite{yao2022controlvae}.
Baumeister et al. employed latent variable mapping for initial control input guidance, where VAE-inferred parameters are fed into neural networks that map latent variables to near-optimal initial control inputs \cite{baumeister2018deep}. 
This approach ensures system stabilization and maintains high objective function values by providing informed starting parameters for MPC optimization.
In both cases, the VAE played a critical role in structuring the search space for MPC optimization, either through a generative latent policy or through initialization guidance, demonstrating how generative modeling can accelerate and stabilize MPC decision-making.

In summary, VAEs distinguish themselves by contributing evenly across all major MPC components: state estimation, predictive modeling, and optimization.
Whether used to infer hidden system states, to approximate complex dynamics, or to structure the optimization search space, the common thread is the use of latent representations.
These representations provide compact and expressive encodings that simplify learning and control, making VAEs a versatile generative tool for enhancing MPC in diverse settings.

Table \ref{tab:ml_integration_mpc} summarizes how GANs, Normalizing Flows, and VAEs integrate with MPC across different components and application domains. 
The table details each method's role in MPC, whether in predictive modeling, state estimation, or optimization, along with its integration strategies and specific applications, demonstrating the diverse approaches these models employ in enhancing MPC capabilities.

%%%%%%%%%%%%%%%%%%%%%%%%%%%%%%%%%%%%%%%%%%%
% TABLE - Review GANs, NFs, VAEs

\begin{sidewaystable*}[p]
  \centering
  \small
  \setlength{\tabcolsep}{5pt}
  \renewcommand{\arraystretch}{1.2}
  \caption{Reviewed Studies on the Integration of GANs, Normalizing Flows, and VAEs with MPC}
  \label{tab:ml_integration_mpc}
  \begin{tabular}{
    >{\raggedright\arraybackslash}p{3.0cm}  % Reference
    >{\raggedright\arraybackslash}p{1.3cm}  % Generative Method
    >{\raggedright\arraybackslash}p{2.3cm}  % Role in MPC
    >{\raggedright\arraybackslash}p{3.4cm}  % Objectives
    >{\raggedright\arraybackslash}p{6.6cm}  % Key Idea
    >{\raggedright\arraybackslash}p{3.0cm}  % Integration Style
    >{\raggedright\arraybackslash}p{2.6cm}  % Application
  }
  \toprule
  \textbf{Reference} & \textbf{Generative Method} & \textbf{Role in MPC} & \textbf{Objectives} & \textbf{Key Idea} & \textbf{Integration Style} & \textbf{Application} \\
  \midrule

  % ===================== GAN =====================
  Li et al.\ (2025) \cite{li2025generative} & GAN & Predictive Modeling & Enable operator foresight in welding &
  Conditional GAN + GRU generates future weld pool images from speed variations &
  Surrogate model (visual prediction) & Welding (GMAW) \\
  Bae et al.\ (2022) \cite{bae2022lane} & GAN & Predictive Modeling & Ensure safe lane changes in dense traffic &
  Social GAN forecasts neighboring vehicles’ trajectories, integrated into MPC collision-avoidance constraints &
  Augmentation (safety-aware scenarios) & Autonomous Driving \\
  Yang et al.\ (2022) \cite{yang2022sms} & GAN & Predictive Modeling & Learn robust robot dynamics &
  Generator learns multi-step dynamics; discriminator enforces realism and temporal consistency &
  Surrogate model (learned dynamics) & Mobile Robots \\
  Burnwal et al.\ (2023) \cite{burnwal2023gan} & GAN & Optimization & Mimic expert behaviors &
  MPC control law adversarially aligned with expert trajectories &
  Control law optimization & Robotics \\
  Gupta et al.\ (2025) \cite{gupta2025adversarial} & GAN & Optimization & Automate reward design &
  MPC expert trajectories train AIRL discriminator to produce reward signals guiding RL optimization &
  MPC-guided RL (reward shaping) & Process Control (Bioreactor) \\

  % ===================== NF =====================
  Cramer et al.\ (2024) \cite{cramer2024least} & Normalizing Flows & Predictive Modeling & Capture stochastic dynamics for robust predictive control &
  Conditional NF models probability distribution of state increments; supports scenario generation and chance constraints &
  Stochastic dynamics modeling & General nonlinear systems \\
  Zhang et al.\ (2024) \cite{zhang2024imflow} & Normalizing Flows & Optimization & Estimate control inputs and quantify uncertainty &
  Conditional NF maps observations to input sequences, enabling probabilistic inverse modeling for MPC &
  Inverse modeling & Vehicle control / trajectory tracking \\
  Power et al.\ (2024) \cite{power2024learning} & Normalizing Flows & Optimization & Improve sampling efficiency and generalization &
  Context-conditioned NF replaces base distribution in MPPI/iCEM, enabling efficient and OOD-robust sampling &
  Sampling distribution learning & Robot navigation / motion planning \\
  Rabenstein et al.\ (2024) \cite{rabenstein2024sampling} & Normalizing Flows & Optimization & Enhance exploration efficiency &
  NF trained offline on heuristic-guided trajectories replaces Gaussian sampler in MPPI &
  Sampling distribution learning & Autonomous driving / trajectory planning \\
  Sacks et al.\ (2023) \cite{sacks2023learning} & Normalizing Flows & Optimization & Accelerate updates and ensure constraint satisfaction &
  NF latent space enables warm-starting, online updates, and automatic constraint handling &
  Latent space optimization & Simulated robotics (navigation, manipulation) \\

  % ===================== VAE =====================
  Yao et al.\ (2022) \cite{yao2022controlvae} & VAE & Predictive Modeling & Approximate unknown dynamics for realistic motion control &
  Learns latent dynamics for physics-based characters, enabling diverse motions via compact encodings &
  Surrogate model (latent dynamics) & Physics-based character animation \\
   &  & Optimization & Enable efficient behavior generation &
  Encodes motion skills as latent variables, used as a generative control policy &
  Latent policy optimization &  \\
  Baumeister et al.\ (2018) \cite{baumeister2018deep} & VAE & State Estimation & Infer hidden variables for robust control &
  VAE infers non-measurable birefringence from laser states &
  Augmentation (soft sensing) & Mode-locked fiber lasers \\
   &  & Optimization & Improve initialization for MPC &
  Maps inferred latent variables to near-optimal initial control inputs &
  Initialization guidance &  \\
  Kwon et al.\ (2020) \cite{kwon2020combining} & VAE & State Estimation & Compact state inference from high-dimensional data &
  Kalman filter embedded in latent space infers compact states from raw images &
  Augmentation (latent state estimation) & Robotic control (air hockey task) \\
   &  & Predictive Modeling & Enable efficient long-term prediction &
  Learns deterministic, time-invariant latent dynamics in low-dimensional state space &
  Surrogate model (latent dynamics) &  \\
  Estiri \& Mirinejad (2024) \cite{estiri2024variational} & VAE & Predictive Modeling & Capture nonlinear physiological dynamics &
  Identifies nonlinear state–output relations and integrates variational Gaussian inference &
  Surrogate model (nonlinear state-space) & Clinical fluid resuscitation \\

  \bottomrule
  \end{tabular}
\end{sidewaystable*}

%%%%%%%%%%%%%%%%%%%%%%%%%%%%%%%%%%%%%%%%%%%

\subsubsection{Roles of Diffusion Models in MPC}

%(Diffusion for Predictive Modeling in MPC)
Diffusion has been integrated into MPC as a predictive mechanism in diverse ways, ranging from trajectory dynamics surrogates to uncertainty forecasting and scenario generation.
Zhou et al. introduced D-MPC, where diffusion was trained as a trajectory-level dynamics model to capture long-horizon and multi-modal behaviors \cite{zhou2024diffusion}.
By learning entire state sequences rather than single-step transitions, their approach reduced compounding errors and enabled more reliable forecasts for downstream planning.
Diffusion served as a stochastic forecasting model in the work of Zarifis et al., where DDPMs combined with recurrent architectures generated uncertainty-aware trajectory predictions for stochastic MPC \cite{zarifis2025diffusion}.
This allowed the controller to evaluate multiple probabilistic futures and improve robustness in energy arbitrage tasks.
Xu and Zhu applied diffusion as a data augmentation method, enriching time series inputs for load forecasting \cite{xu2025diffusion}.
The diffusion-assisted framework enhanced dataset diversity and representativeness, thereby improving the accuracy of the load forecasting model and strengthening MPC performance in power system operations.
Diffusion was leveraged as a multi-agent scenario generator in Samavi et al., where the diffusion framework produced diverse human and robot trajectories for safe crowd navigation \cite{samavi2024safe}.
The generated trajectories were integrated into a bilevel MPC structure with explicit safety constraints, ensuring robust collision avoidance in dense and uncertain environments.
%As shown in the studies above, diffusion in predictive roles supports MPC by modeling long-horizon system dynamics, generating uncertainty-aware forecasts, augmenting data to improve input quality, and producing multi-agent scenarios for safety-aware control.

%(Diffusion for Optimization in MPC)
Diffusion models have also been leveraged in MPC optimization through policy learning and imitation, constraint handling, and global optimization.
% Diffusion - Optimization - Policy Learning
In policy learning and imitation, diffusion has been used to approximate or guide MPC policies by learning from demonstrations or shaping action proposals. 
Zhou et al. applied diffusion to learn action sequence proposal distributions, which accelerated trajectory search and allowed MPC to evaluate diverse candidate solutions more efficiently \cite{zhou2024diffusion}. 
Julbe extended this approach to direct policy imitation, training diffusion models on expert MPC trajectories to capture multi-modal control distributions that conventional least-squares models cannot represent, thus making real-time deployment feasible \cite{julbediffusion}. 
Building on this work, Julbe et al. introduced gradient guidance and early stopping into the diffusion process, stabilizing policy imitation and enabling high-frequency execution without unstable mode switching \cite{julbe2025diffusion}.
% Diffusion - Optimization - Handle constraints
Diffusion has also been integrated to handle constraints during optimization. 
Römer et al. introduced iterative model-based projection into the backward denoising process, applying constraint tightening to account for model mismatch and generating trajectories that satisfy dynamic feasibility even for novel constraints \cite{romer2024diffusion}.
Samavi et al. applied diffusion within a bilevel MPC framework for crowd navigation, where diffusion-generated human trajectory samples were filtered at a lower level to exclude unsafe motions, ensuring that the final MPC plan remained collision-free \cite{samavi2024safe}.
% Diffusion - Optimization - Global Optimization
Finally, global optimization is also a field where diffusion methods are applied. 
Huang et al. modeled the distribution of local NMPC solutions using diffusion, which enabled MPC to perform sample–score–rank optimization across multiple solution modes. 
This approach extended the search beyond local minima, providing a probabilistic pathway toward near-global optimality \cite{huang2024toward}.
%Diffusion-based optimization enhances MPC by enabling fast and flexible policy learning, embedding constraint satisfaction directly into the generation process, and extending optimization toward more globally optimal solutions in nonlinear control problems.

In summary, diffusion models have emerged as versatile generative tools for MPC, contributing both to predictive modeling and to optimization. 
For predictive modeling, diffusion enables trajectory-level forecasting, stochastic scenario generation, uncertainty-aware prediction, and data augmentation, allowing MPC to better anticipate complex, multimodal system behaviors. 
For optimization, diffusion supports policy learning and imitation, embeds feasibility and safety constraints directly into the generation process, and facilitates near-global exploration of nonlinear control problems. 
Together, these advances highlight the ability of diffusion models to enhance both the foresight and the decision-making capacity of MPC, extending its applicability to highly uncertain, multimodal, and safety-critical environments.

\subsubsection{Roles of LLMs in MPC}
LLMs have recently been applied to MPC for supporting tasks such as state estimation, predictive modeling, and optimization. 
In parallel, Vision-Language Models (VLMs) have emerged as variations of LLMs that integrate a vision encoder to process images or video alongside text. 
They extend the language modeling capability of LLMs by grounding it in visual perception, enabling tasks that require joint use of visual and textual information \cite{zhang2024vision, bordes2024introduction}.

%(LLMs and VLMs for Predictive Modeling in MPC)
LLMs and VLMs contribute to predictive modeling in MPC, particularly by capturing sequential dynamics and enabling adaptive forecasting.
One study demonstrated how an autoregressive language model can serve as a transition model for human–robot interaction, generating multiple rollouts of future trajectories and allowing MPC to select the shortest path to task completion \cite{liang2024learning}.
Another work applied a language model as a topology-adaptive surrogate for power systems, where in-context learning enabled accurate voltage prediction under frequent topology changes with minimal data requirements \cite{jena2025llm}.
On the perceptual side, a vision–language model was integrated into robotic manipulation tasks, where action-conditioned video prediction provided foresight of future frames, improving MPC’s ability to plan toward sub-goals while accounting for potential interferences \cite{zhao2024vlmpc}.

%(LLMs and VLMs for Optimization in MPC)
LLM- and VLM-based optimization enhance MPC via three roles.
% LLM generate plan
The first is plan proposal and selection, where LLMs act as approximate optimizers by generating candidate action sequences that are then evaluated by explicit cost functions. 
Maher illustrated this by using an LLM to generate various candidate plans, simulate their results, and choose the one with the lowest cost, effectively replacing the optimization loop with model-generated proposals \cite{maher2025llmpc}.
The second role is high-level task guidance with safety checks, in which VLMs or LVLMs generate symbolic task commands (e.g., lane change, stop, and accelerate) while MPC verifies their feasibility and ensures safe execution. 
Atsuta et al. \cite{atsuta2025lvlm} presented such a collaborative architecture for autonomous driving, where the LVLM provided high-level driving maneuvers and the MPC builder validated and executed them while handling transitions between feasible and infeasible tasks.
The third role is vision-guided trajectory optimization, which leverages VLMs’ perception capabilities to inform cost functions and trajectory search. 
Zhao et al. \cite{zhao2024vlmpc} incorporated VLM-assisted cost functions into MPC for robotic manipulation, embedding knowledge of sub-goals and obstacles into the optimization process. 
Similarly, Chen et al. \cite{chen2025vision} employed VLMs to generate 3D trajectory hypotheses and voxel-based value maps, enabling efficient evaluation of motion plans for long-horizon tasks.

In summary, LLMs and VLMs extend MPC across state estimation, predictive modeling, and optimization by bridging high-dimensional information with control objectives.
They provide contextual state inference, model sequential dynamics, and guide optimization through plan generation, symbolic task commands, or perception-informed cost shaping, highlighting their versatility as flexible interfaces that connect complex data to structured decision-making.

Table \ref{tab:ml_integration_mpc_2} presents the integration of Diffusion Models, LLMs, and VLMs in MPC systems, showing their roles across MPC components, key implementation strategies, and application domains. These methods demonstrate distinct integration approaches, from diffusion-based trajectory generation to language model-guided planning and vision-language multimodal understanding.

%%%%%%%%%%%%%%%%%%%%%%%%%%%%%%%%
% TABLE - Gen ML-driven MPC - Diffusion, LLM, VLM %

\begin{sidewaystable*}[p]
  \centering
  \small
  \setlength{\tabcolsep}{5pt}
  \renewcommand{\arraystretch}{1.2}
  \caption{Reviewed Studies on the Integration of Diffusion Models, LLM, and VLMs with MPC}
  \label{tab:ml_integration_mpc_2}
  \begin{tabular}{
    >{\raggedright\arraybackslash}p{3.0cm}  % Reference
    >{\raggedright\arraybackslash}p{1.3cm}  % Generative Method
    >{\raggedright\arraybackslash}p{2.3cm}  % Role in MPC
    >{\raggedright\arraybackslash}p{3.4cm}  % Objectives
    >{\raggedright\arraybackslash}p{6.6cm}  % Key Idea
    >{\raggedright\arraybackslash}p{3.0cm}  % Integration Style
    >{\raggedright\arraybackslash}p{2.6cm}  % Application
  }
 \toprule
  \textbf{Reference} & \textbf{Generative Method} & \textbf{Role in MPC} & \textbf{Objectives} & \textbf{Key Idea} & \textbf{Integration Style} & \textbf{Application} \\
  \midrule

  % ===================== Diffusion =====================
  Zhou et al.\ (2024) \cite{zhou2024diffusion} & Diffusion & Predictive Modeling & Capture long-horizon, multimodal dynamics &
  Diffusion trained as trajectory-level surrogate model, reducing compounding errors &
  Surrogate model (dynamics) & Continuous control (D4RL) \\
   &  & Optimization & Accelerate trajectory search &
  Learns action sequence proposal distributions for stochastic planners &
  Action proposal learning &  \\
  Julbe (2025) \cite{julbediffusion} & Diffusion & Optimization & Approximate MPC policy &
  Trains diffusion on MPC expert trajectories to capture multimodal policies for real-time use &
  Policy imitation (offline) & Robotic arm control (7-DOF KUKA) \\
  Julbe et al.\ (2025) \cite{julbe2025diffusion} & Diffusion & Optimization & Stabilize high-frequency MPC policy &
  Adds gradient guidance and early stopping in denoising for stable, high-frequency imitation &
  Policy imitation (stabilized online) & Robotic manipulation (7-DOF KUKA) \\
  Zarifis et al.\ (2025) \cite{zarifis2025diffusion} & Diffusion & Predictive Modeling & Provide uncertainty-aware forecasts &
  Combines diffusion with recurrent models for stochastic trajectory prediction &
  Surrogate model (stochastic forecasting) & Energy arbitrage (battery storage) \\
  Xu \& Zhu (2025) \cite{xu2025diffusion} & Diffusion & Predictive Modeling & Improve input data quality &
  Uses diffusion for data augmentation to enrich load forecasting datasets &
  Data augmentation & Power system operations (IEEE 30-bus) \\
  Samavi et al.\ (2024) \cite{samavi2024safe} & Diffusion & Predictive Modeling & Generate safe multi-agent scenarios &
  Diffusion generates diverse human–robot trajectories integrated into bilevel MPC &
  Scenario generation (multi-agent) & Crowd navigation \\
   &  & Optimization & Enforce safety constraints &
  Unsafe trajectories filtered before MPC planning &
  Constraint filtering &  \\
  Römer et al.\ (2024) \cite{romer2024diffusion} & Diffusion & Optimization & Ensure dynamic feasibility &
  Adds iterative projection into denoising to enforce feasibility under model mismatch &
  Constraint projection & Robotic manipulator simulation \\
  Huang et al.\ (2024) \cite{huang2024toward} & Diffusion & Optimization & Achieve near-global optimality &
  Models distribution of local NMPC solutions to explore multimodal solution space &
  Global optimization & Nonlinear control (Cart-pole, Pendubot) \\

  % ===================== LLM =====================
  Liang et al.\ (2024) \cite{liang2024learning} & LLM & Predictive Modeling & Model sequential HRI dynamics &
  LLM autoregressively predicts interaction rollouts; MPC selects shortest path &
  Surrogate model (interaction dynamics) & Human–Robot Interaction \\
  Jena et al.\ (2025) \cite{jena2025llm} & LLM & Predictive Modeling & Provide topology-adaptive surrogate &
  LLM approximates power flow for voltage prediction, adapting via in-context learning &
  Surrogate model (power flow) & Power Systems Control \\
  Maher (2025) \cite{maher2025llmpc} & LLM & Optimization & Propose and refine control sequences &
  LLM generates candidate action plans; cost functions evaluate and select best &
  Plan proposal \& evaluation & General Planning \\

  % ===================== VLM =====================
  Long et al.\ (2024) \cite{long2024vlm} & VLM & State Estimation & Enhance perception of environment and states &
  VLM encodes environment and vehicle states, supported by reference memory &
  MPC parameter generation & Autonomous Driving (nuScenes) \\
   &  & Optimization & Improve planning stability and reliability &
  Upper-layer planner generates MPC parameters; lower-layer MPC executes in real time &
  MPC parameter tuning &  \\
  Zhao et al.\ (2024) \cite{zhao2024vlmpc} & VLM & Predictive Modeling & Anticipate visual outcomes of actions &
  VLM predicts action-conditioned video frames to foresee future states &
  Hybrid prediction & Robotic Manipulation \\
   &  & Optimization & Guide planning with visual reasoning and constraints &
  VLM provides sub-goals and knowledge-level constraints for cost definition and sampling &
  Hybrid (prediction + constraints) &  \\
  Atsuta et al.\ (2025) \cite{atsuta2025lvlm} & VLM & Optimization & Scale task planning and ensure safety &
  LVLM generates symbolic task plans; MPC checks feasibility and synthesizes iOCPs &
  Task-to-control interface & Autonomous Driving (Highway) \\
  Chen et al.\ (2025) \cite{chen2025vision} & VLM & Optimization & Generate efficient long-horizon trajectories &
  VLM-driven GMM samples 3D trajectories; voxel-based maps assess feasibility &
  Trajectory sampling \& evaluation & Robotic Manipulation (Long-Horizon) \\

  \bottomrule
  \end{tabular}
\end{sidewaystable*}

%%%%%%%%%%%%%%%%%%%%%%%%%%%%%%%%

%Explain how generative ML enhances the principle of each component of MPC with mathematics and show application examples
\subsubsection{Synthesis: Generative ML Integration Across MPC Components}

In this subsection, we synthesize how generative ML transforms each major MPC component and analyze its applications in manufacturing processes. 

For predictive modeling, generative ML has been primarily applied as a surrogate system model, as shown in our review. 
Traditional MPC relies on analytical system models that capture real-world dynamics through simplified mathematical representations. 
In the system modeling formulation given in Equation \ref{eq:MPC_system_formulation}, where $f$ represents the system dynamics, this analytical function can be replaced by a generative surrogate $f_{\text{GenML}}$. 
Depending on the chosen generative family, the surrogate takes different mathematical forms, as formalized in the representative formulations of GANs (Equation~\ref{Eq:gan_objective}), NFs (Equation~\ref{eq:nf_changevar}), VAEs (Equation~\ref{eq:vae_encoder}), diffusion (Equation~\ref{Eq:diff_reverse}), and LLMs (Equation~\ref{eq:llm_autoreg}). 
These models thus extend predictive modeling beyond traditional analytical surrogates.

Generative surrogates can be expressed in both deterministic and probabilistic forms, as shown in Equations \ref{eq:genml_prob_1} and \ref{eq:genml_prob_2}, respectively.
\begin{align}
\hat{x}_{k+1} &= f_{\text{GenML}}(x_k, u_k) \label{eq:genml_prob_1}\\
p(\hat{x}_{k+1}\mid x_k, u_k) &= f_{\text{GenML}}(x_k, u_k)
\label{eq:genml_prob_2}
\end{align}

This transformation enables three fundamental enhancements that transcend the limitations of traditional analytical modeling. 
\begin{enumerate}
     
\item\textbf{Enhanced System Dynamics Learning.} 
Generative ML models learn complex nonlinear dynamics directly from data, capturing intricate relationships that analytical models cannot represent. 
VAEs learn compressed latent dynamics \cite{yao2022controlvae, kwon2020combining}, GANs generate realistic state transitions through adversarial training \cite{li2025generative,bae2022lane,yang2022sms}, and diffusion models capture complex trajectory-level behaviors \cite{zhou2024diffusion,samavi2024safe}. 
This data-driven approach enables MPC to handle manufacturing processes with unknown or highly complex dynamics that resist analytical formulation. 

\item\textbf{Multi-mode Distribution Modeling.} 
Unlike deterministic analytical models that produce single-valued predictions, generative ML naturally represents uncertainty and multiple possible outcomes through probabilistic distributions with multiple modes \cite{bae2022lane,julbe2025diffusion,yao2022controlvae}. 
This capability proves essential for manufacturing processes exhibiting stochastic behaviors, multiple operational modes, or process variations. 
The probabilistic formulation, which is Equation \ref{eq:genml_prob_2}, enables robust decision-making under uncertainty, addressing one of the fundamental limitations of traditional MPC. 

\item\textbf{Trajectory-level Representation.} 
Generative ML enables learning of entire trajectory distributions rather than single-step transitions, addressing the compounding error problem inherent in recursive analytical models. 
Diffusion models excel at capturing joint trajectory distributions, while sequence models like LLMs naturally handle temporal dependencies across extended horizons\cite{zhou2024diffusion,liang2024learning}. 
This capability proves particularly valuable for manufacturing processes requiring long-horizon planning and coordination. 

\end{enumerate}

In manufacturing, generative surrogates have been employed to improve predictive modeling in several ways. 
For instance, conditional GANs have been used to generate weld pool images conditioned on torch speed, enabling human operators to visualize future states before execution in welding-based MPC \cite{li2025generative}. 
Similarly, Social GANs have supported predictive safety constraints in autonomous driving scenarios, and diffusion-based approaches have been explored for multi-agent trajectory forecasting and stochastic load prediction, both of which parallel manufacturing challenges in multi-robot coordination and energy system control~\cite{bae2022lane,samavi2024safe,xu2025diffusion}. 
These applications demonstrate how generative predictive models can enhance MPC’s ability to anticipate process behavior under uncertainty and complex interactions, which are central to advanced manufacturing environments.

% State Estimation
MPC relies on state estimators to infer system states when they cannot be directly measured or when sensor data is noisy \cite{Schwenzer21}. 
Traditional approaches such as Kalman filters assume linear dynamics, Gaussian noise, and explicit mathematical models of process and measurement uncertainties \cite{wu2025tutorial,kwon2020combining}. 
Manufacturing environments often violate these assumptions, presenting high-dimensional sensor inputs, non-Gaussian uncertainties, and unmeasurable process variables \cite{gunasegaram2024machine,zheng2023physics}. 
Generative ML enhances state estimation by learning data-driven latent representations that capture hidden or complex variables without requiring explicit models of uncertainty \cite{baumeister2018deep,kwon2020combining}. 
The reviewed studies highlight three complementary integration patterns.
\begin{enumerate}
\item \textbf{Latent Variable Inference.}  
VAEs demonstrate effectiveness in learning probabilistic mappings between observable quantities and hidden variables, allowing MPC to understand the current state and maintain control performance despite uncertainty\cite{baumeister2018deep,kwon2020combining}.
This preprocessing approach transforms unmeasurable states into observable quantities that can then be processed by later MPC stages.

\item \textbf{High-Dimensional Sensor Processing.}  
Manufacturing systems increasingly rely on high-dimensional sensor data that exceeds the processing capabilities of traditional state estimators. Generative ML addresses this challenge by compressing complex sensory information into lower-dimensional state representations suitable for MPC processing.

\item \textbf{Multimodal State Encoding.}  
Complex manufacturing environments require integration of diverse sensor modalities with different characteristics \cite{chen2024multi}. 
In our reviewed study, VLMs enable sophisticated preprocessing by combining visual understanding with the extraction of unified state representations from heterogeneous sensor streams \cite{long2024vlm}. 
This capability enables MPC to operate in scenarios where environmental context and historical patterns are essential for accurate state estimation.
\end{enumerate}

In manufacturing, these advances allow MPC to operate effectively in challenging settings: VAEs have been used to infer birefringence in optical systems, robotic control benefits from vision-based latent encodings, and autonomous manufacturing vehicles leverage VLMs to contextualize complex environments for safe operation \cite{baumeister2018deep,kwon2020combining,long2024vlm}.

%(Optimization)
MPC optimization often uses sampling-based solvers or random sampling that often get stuck in local optima and struggle with complex constraints \cite{wu2025tutorial,Schwenzer21}. Generative ML has been shown to transform this core process by introducing three practical approaches that improve how MPC finds control solutions.
\begin{enumerate}
     
\item \textbf{Multiple Solution Exploration.} 
Manufacturing systems often have multiple good operating strategies. Diffusion models learn to sample diverse trajectory options instead of converging to single solutions \cite{huang2024toward, zhou2024diffusion}. 
NFs replace naive Gaussian sampling with learned distributions that focus on high-quality regions \cite{power2024learning,rabenstein2024sampling}. 
This capability of generative MLs prevents MPC from getting trapped in suboptimal local solutions \cite{huang2024toward}.

\item \textbf{Constraint-Aware Generation.}
Traditional optimization handles constraints through a cost function or hard constraints. 
Generative ML builds constraint satisfaction directly into the solution generation process. 
This approach is particularly valuable for manufacturing systems with complex safety operational limitations.
Diffusion models use iterative refinement to generate trajectories that naturally satisfy safety and collision avoidance constraints \cite{samavi2024safe, romer2024diffusion}. 
NFs integrate control limits into their architecture by design \cite{sacks2023learning}. 

\item \textbf{Learned Control Strategies.}
Generative ML learns control policies from demonstrations or implicit objectives. 
GANs train MPC policies through adversarial learning, mimicking expert behavior without explicit reward engineering \cite{burnwal2023gan, gupta2025adversarial}. 
LLMs generate diverse control plans, providing high-level commands that guide MPC execution  \cite{maher2025llmpc, atsuta2025lvlm}. 
This learned approach handles manufacturing tasks where optimal behavior is easier to demonstrate than to specify mathematically \cite{gupta2025adversarial}.
\end{enumerate}

These optimization capabilities support broader manufacturing control strategies beyond current applications. 
Multiple solution exploration supports adaptive manufacturing, where systems must switch between different operational modes based on product variants or environmental conditions \cite{huang2024toward}. 
Constraint-aware generation enables safe operation in shared human-robot or multi-robot workspaces \cite{samavi2024safe}. 
Learned control strategies allow manufacturing systems to adapt to new products or processes through demonstration rather than extensive reprogramming, particularly valuable for flexible manufacturing environments and custom production scenarios \cite{burnwal2023gan, gupta2025adversarial}.

\begin{figure*}[thbp]
    \centering
    \includegraphics[width=\textwidth]{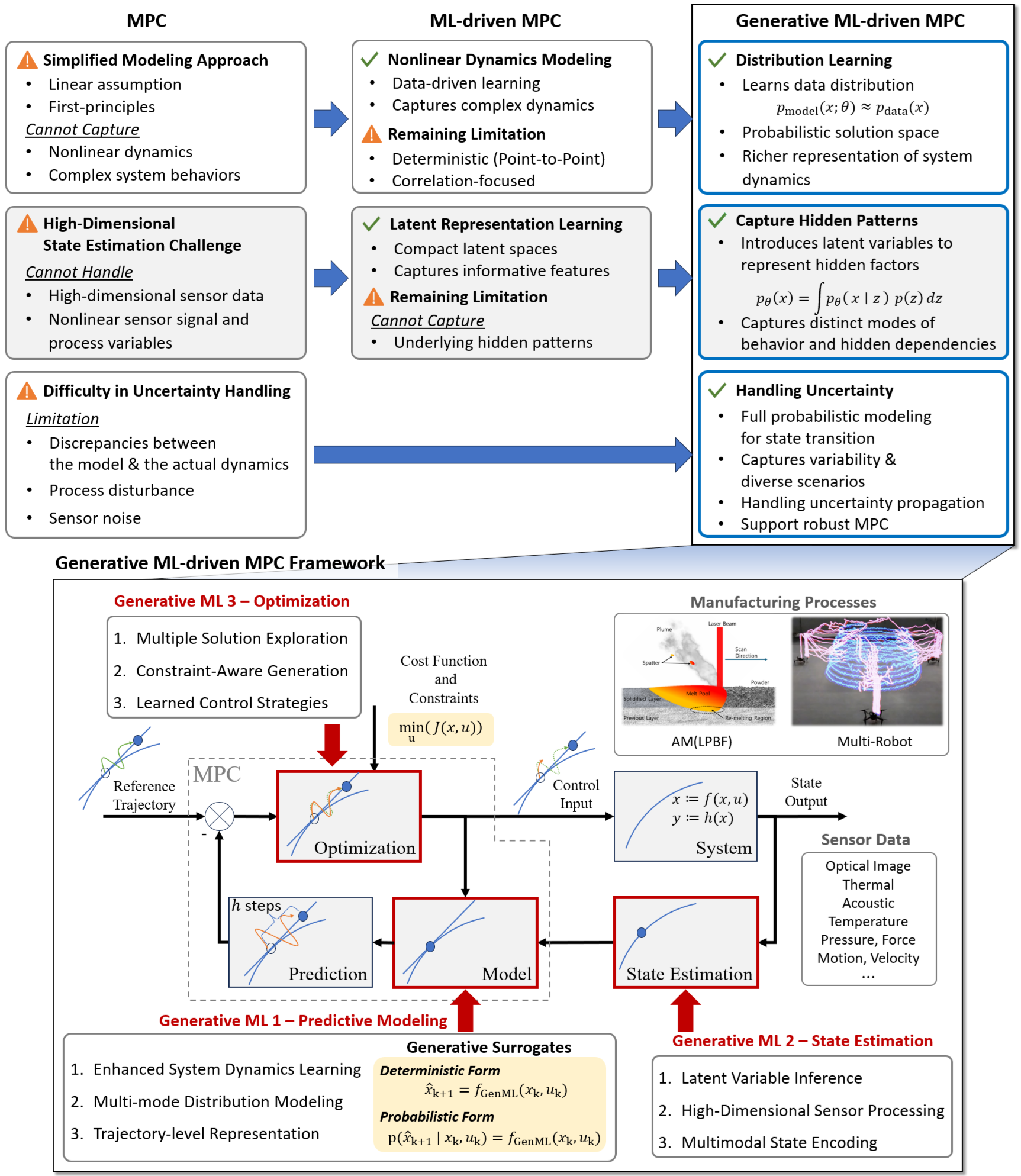}
    \caption{Progression from MPC to generative ML-driven MPC and the resulting integrated framework. The upper panel shows how conventional MPC, ML-driven MPC, and generative ML-driven MPC compare, highlighting how generative approaches progressively address limitations of previous approaches.
    The lower panel depicts the generative ML-driven MPC framework, where generative ML augments each core MPC component. 
    In predictive modeling, generative surrogates replace conventional models, offering both deterministic and probabilistic formulations.
    In state estimation, generative ML enables latent variable inference, multimodal sensor encoding, and high-dimensional data processing.
    In optimization, it supports multiple-solution exploration, constraint-aware generation, and learned control strategies.
    Representative manufacturing processes (e.g., LPBF \cite{lee2024amtransformer} and multi-robot systems \cite{zhang2022aerial}) and diverse sensor modalities are also shown as application contexts, emphasizing the practical relevance of generative ML-driven MPC.}

    \label{fig:genml_mpc_framework}
\end{figure*}

To summarize, the integration of generative ML with MPC extends beyond individual methods and provides a comprehensive framework. 
Figure~\ref{fig:genml_mpc_framework} presents an overview of how generative ML transforms each major MPC component, illustrating its contributions to state estimation, predictive modeling, and optimization.

%%%%% Figure %%%%%

%%%%%%%%%%%%%%%%%%%%%%%%%%%%%%%%%%%%%%%%%%%%%%%%%%%%%%%%%%%%%%%%%%%%%%%%%%%%%%%
\section{Research Gaps}

While generative ML-driven MPC has demonstrated promising potential across diverse application domains, several limitations remain before these methods can be reliably deployed in real manufacturing environments. 
Our review reveals three critical areas where current approaches fall short: (1) the limited application to manufacturing settings and the lack of integration of domain-specific knowledge, (2) inadequate robustness when exposed to distribution shifts or unseen operating conditions, and (3) inference latency that conflicts with real-time control requirements of MPC.
Together, these gaps illustrate the current limitations that prevent generative ML-driven MPC from being applied in manufacturing processes.

%Limited application to manufacturing, lack of integration of domain knowledge
Despite the growing interest in generative ML-driven MPC, its exploration in manufacturing research remains limited. 
Most existing studies have focused on domains such as robotics and autonomous driving, with only a handful directly addressing manufacturing processes. 
While ML-driven MPC has already been actively explored in manufacturing research, generative ML-driven MPC remains scarce, resulting in a limited understanding of how it performs under manufacturing processes. 
Moreover, even in the few manufacturing-related studies that exist, generative models were primarily employed in data-driven roles. 
While such models implicitly reflect some constraints and system behaviors through training data, they did not explicitly encode manufacturing-specific knowledge, such as process physics, operational limits, or quality requirements, in their architectures or objectives. 
This lack of explicit knowledge integration restricts the ability of current approaches to demonstrate feasibility and reliability in manufacturing process control.

%inadequate robustness when exposed to distribution shifts
Another research gap concerns the limited robustness of generative ML-driven MPC when exposed to distribution shifts or unseen operating conditions. 
Most existing studies have evaluated their methods within training-like conditions, without systematically testing performance under variations that are common in manufacturing, such as parameter drift, material heterogeneity, or equipment variability. 
Although a few isolated attempts have explored sampling strategies for improved generalization, comprehensive frameworks for out-of-distribution robustness remain absent. 
This lack of systematic assessment makes it difficult to establish whether generative ML-driven MPC can reliably maintain control performance when manufacturing conditions deviate from the training distribution. 
%\textcolor{blue}{RS: Is this something that is manufacturing-specific? This seems like more of a general problem of generative ML-driven MPC. Are there other domains in which this paradigm has been successfully applied under distribution shifts?}

%Inference latency that conflicts with real-time control requirements of MPC.
A further gap arises from the trade-off between computational complexity and the real-time requirements of MPC. 
Generative ML methods such as diffusion models and LLMs often involve iterative sampling or sequential inference steps that introduce significant latency, making them difficult to deploy in manufacturing environments where control updates must occur on the order of milliseconds. 
Although some studies have proposed early-stopping strategies to mitigate this burden, these efforts have largely been tested in simulation rather than in real manufacturing systems.
As a result, the feasibility of using computationally intensive generative models within the tight timing constraints of manufacturing MPC remains insufficiently examined.

These gaps underscore that generative ML-driven MPC is not yet ready for reliable deployment in manufacturing contexts. 
They highlight the need for further research to establish domain-specific integration, robustness, and real-time feasibility, setting the stage for the future directions discussed in the next section.

%%%%%%%%%%%%%%%%%%%%%%%%%%%%%%%%%%%%%%%%%%%%%%%%%%%%%%%%%%%%%%%%%%%%%%
\section{Future Research}

Based on the identified research gaps, this section outlines promising directions for future work and presents application examples that illustrate how generative ML can potentially address the main limitations of MPC, shown here through a representative multi-robot manufacturing case but broadly applicable to other domains.

\subsection{Research Direction}
% Manufacturing Knowledge-Integrated Generative Architectures
%Propose embedding manufacturing-specific constraints into generative ML architectures using approaches such as PINNs.
First, future research should focus on integrating manufacturing knowledge into generative ML-driven MPC architectures. 
In the studies reviewed, generative models were able to learn data distributions, but these distributions reflected only the statistical patterns in the training data and did not fully capture specific manufacturing requirements. 
While the MPC optimizer naturally handles constraints such as control limits or safety conditions, generative models that lack embedded domain knowledge may still produce proposals that are infeasible or misaligned with process physics and quality requirements. 
Future work should therefore investigate ways to enrich these learned distributions with manufacturing-specific knowledge so that the generated candidates are not only data-consistent but also physically meaningful before optimization takes place. 
For example, process physics can be incorporated into the training objective, operational limits can be explicitly enforced during generation, and quality requirements can be represented as conditions within the model. 
By enhancing the learned distributions in this way, generative ML-driven MPC would move beyond reproducing statistical patterns and provide outputs that are both practically feasible and reliable in real manufacturing settings. %\textcolor{blue}{RS: Would the use of negative data, highlighting infeasible, non-physical solutions, also be beneficial here? Check this out if there is time: https://openreview.net/forum?id=FNBv2vweBI} 
At the same time, embedding physical and operational constraints into the generative models would also improve explainability, since the generated outputs could be interpreted with respect to known process laws and quality requirements rather than only as statistical reproductions of past data.

% Foundation Model-based Approaches
% Propose developing foundation model-based approaches pretrained on manufacturing data and fine-tuned for specific applications to improve adaptability.

Second, future research should address the lack of robustness under distribution shifts. 
Most of the studies reviewed evaluated generative ML-driven MPC only under the similar conditions as the training data, with little evidence on how these methods perform when process conditions change or when unseen disturbances arise. 
This limitation is critical for manufacturing, where variations in material properties, equipment conditions, or operating environments are inevitable. 
For example, in semiconductor fabrication, even when two machines are of the same model, their calibration differences or setup variations can significantly alter process behavior, creating distribution shifts that challenge model robustness. 
Future work should therefore establish systematic evaluation protocols that explicitly test performance under out-of-distribution scenarios such as unseen disturbances or process variations. 
A promising direction is the development of foundation models trained on diverse manufacturing datasets that capture a wide range of scenarios. These models could then be adapted to specific processes through fine-tuning, thereby improving both generalization and reliability in real-world manufacturing environments. 
%\textcolor{blue}{RS: Are you suggesting some form of transfer learning?}.

% Applying advanced approaches to the model in MPC to accelerate the inference time. Adjust of appropriate level of abstraction of modeling that can achieve computation speed and performance.
Third, future research should address the latency of generative ML models, which currently limits their use in real-time manufacturing MPC. 
Generative ML methods, such as diffusion models, LLMs, and VLMs, often require repeated sampling or sequential inference steps, which introduce latency that conflicts with high-frequency control updates in manufacturing, often down to the millisecond scale. 
Some prior works have used early-stopping or truncated iterations to reduce latency, but these are mostly validated in simulation and may not guarantee acceptable accuracy in real manufacturing systems. 
To move forward, research should also take advantage of ongoing advances from the computer science and ML community, which continuously seek to accelerate inference in generative models. Techniques emerging from this field of work could reduce latency while maintaining the accuracy and stability required for integration into real-time manufacturing MPC. 
%\textcolor{blue}{RS: It might be useful to provide a benchmark, or what level of latency would be acceptable for manufacturing. I am unsure whether it is reasonable to expect these models to finish inference in the order of a few milliseconds.}

\subsection{Illustrative Case: Generative ML-driven MPC in Multi-Robot Manufacturing}
% Future Application Scenarios
%Illustrate how generative ML-driven MPC can be applied to multi-robot manufacturing.

% Intro
%\textcolor{blue}{RS: I feel like this section may need to be refined. The problems and the solutions outlined here feel a bit too general, almost like a repeat of 6.1. We may want to give some concrete information on the examples of Figure 5.} Among various future directions, multi-robot manufacturing provides a representative scenario to illustrate the potential of generative ML-driven MPC. 
Multi-robot systems embody the complexity of multi-process manufacturing, which requires tight coordination, operates under significant interaction uncertainties, and relies on various sensing and communication \cite{zhang2022aerial, stone2025safezone, silver2005cooperative, poudel2020heuristic, mensch2024real}.
This makes them a compelling testbed to highlight how generative modeling can augment predictive control under real-world manufacturing conditions. 
To this end, we reviewed MPC-based studies in multi-robot manufacturing and identified recurring limitations, particularly in terms of model uncertainty, imperfect sensing, and the difficulty of controller design and tuning. 
%\textcolor{blue}{RS: We may want to link these difficulties to the three gaps we recently proposed.}
Building on these insights, we illustrate how generative ML can address these challenges and enable more robust and adaptive MPC in multi-robot scenarios. 
These scenarios are not intended to suggest that multi-robot manufacturing is the sole trajectory for future applications, but rather to serve as an illustrative example of how generative ML can reshape predictive control in complex, highly interactive environments.

%Model Uncertainty and Robustness
A major limitation of MPC in multi-robot systems lies in its vulnerability to model mismatch and the difficulty of ensuring robustness under uncertainty. 
Since MPC relies on predictive models of system dynamics, any discrepancy between the model and the actual plant leads to degraded performance, particularly in highly nonlinear and uncertain environments. 
In space-based multi-robot manufacturing, the system faces pronounced robustness challenges due to extreme environmental uncertainties. 
The absence of gravity, variations in stiffness and damping, and uncertain payload parameters such as mass and inertia all exacerbate the mismatch between predictive models and actual system dynamics; bias correction is often required to partially compensate for such effects \cite{kalaycioglu2023passivity}. 
Beyond parameter uncertainty, ensuring stability under nonlinear and complex robot motions remains difficult. 
Much of the MPC literature has focused on linearized models, which are unsuitable for highly nonlinear multi-robot dynamics, while nonlinear MPC does not always guarantee closed-loop stability and is often confined to simplified settings \cite{kalaycioglu2023passivity}. 
Similar limitations appear in terrestrial scenarios; cooperative transport approaches frequently neglect dynamic effects, which constrain performance in highly dynamic environments \cite{muhammed2024multi}; and nonlinear MPC, though effective for trajectory tracking, is rarely deployed in force-sensitive cooperative tasks where precise interaction modeling is required \cite{zhang2022multi}. 
Collectively, these findings highlight that unmodeled dynamics, parameter variability, and nonlinear complexity remain critical bottlenecks for MPC, directly undermining its reliability and safety in multi-robot manufacturing.

%Imperfect Sensing and Perception.
Another critical limitation arises from the imperfection of sensing and the subsequent perception process, as MPC performance depends on accurate and timely state information. 
In practice, these weights often require extensive offline adjustment to achieve acceptable performance in specific tasks \cite{kalaycioglu2023passivity}. 
For instance, iGPS requires an unobstructed line of sight to multiple transmitters, and its performance deteriorates under occlusion or workspace clutter \cite{storm2021state}. 
Vision-based control methods exhibit similar vulnerabilities: although the camera may provide raw images, the perception of object position or depth from these images is strongly affected by lighting conditions and viewing geometry \cite{muhammed2024multi}. 
Even in advanced visual approaches, depth information inferred from two-dimensional images can be inaccurate, while single-view depth cameras often fail to provide reliable distance cues under occlusion or poor illumination \cite{deng2025multi}. 
These limitations stem from both imperfect sensing and environment-dependent perception, which degrade state estimation and reduce the reliability of predictive control in multi-robot manufacturing, where precise and synchronized coordination is essential.

%Controller Design and Tuning Difficulty.
A further limitation of MPC in multi-robot systems is the difficulty of controller design and parameter tuning. 
MPC relies on weighting matrices in its cost function to balance competing objectives such as tracking accuracy, control effort, and smoothness of input signals \cite{kalaycioglu2023passivity}. 
In practice, these weights are often chosen manually and require extensive offline adjustment to achieve acceptable performance in specific tasks \cite{kalaycioglu2023passivity}. 
For example, even advanced nonlinear MPC frameworks still depend on positive definite weighting matrices that must be tuned to the application context \cite{kalaycioglu2023passivity}. 
Cooperative transport tasks similarly define cost functions with multiple weight terms that require careful calibration, while visual servoing approaches note the need for more sophisticated or task-specific mechanisms to set weighting parameters effectively \cite{muhammed2024multi, deng2025multi}. 
This dependence on manual parameterization not only increases design effort but also limits generalizability, as tuning that works in one scenario may fail when task conditions or robot configurations change. 
In multi-robot manufacturing, where coordination tasks vary widely in scale and dynamics, this lack of systematic and adaptive tuning poses a significant barrier to deploying MPC reliably across diverse settings.

%Generative ML-driven MPC is a Solution.
The limitations identified above can be systematically mitigated through the integration of generative ML with MPC. 

%Addressing Model Uncertainty with Generative ML.
When environmental factors such as gravity absence, variations in stiffness or damping, or payload variability introduce uncertainty, generative models could learn the distribution of system responses conditioned on these changing factors \cite{kalaycioglu2023passivity, muhammed2024multi, zhang2022multi}. 
Instead of relying on a single nominal prediction, the controller could explore multiple trajectories sampled from the learned distribution, allowing it to anticipate a range of possible outcomes. 
A similar approach could apply to nonlinear and complex motions of multiple robots \cite{kalaycioglu2023passivity, zhang2022multi}. 
Rather than depending on linearized approximations or simplified assumptions, generative models could capture diverse patterns of nonlinear dynamics and provide multiple candidate plans that reflect this variability. By evaluating and adapting among these alternatives, MPC could more reliably handle the uncertainty and complexity of multi-robot coordination.

For instance, in dual-arm cooperative operations where synchronization mismatch between robots can cause workpiece deformation \cite{zhang2022multi}, generative models could learn the distribution of coupled force interactions that are difficult to capture with simplified analytical models. 
By generating multiple synchronized motion plans and selecting trajectories that account for these interaction dynamics, the approach could reduce deformation risk more effectively than methods relying on decoupled or linearized assumptions.

%Addressing Imperfect Sensing and Perception with Generative ML.
When sensing is degraded by noise or varying environmental conditions, generative models could be used to reconstruct missing or unreliable data \cite{storm2021state, deng2025multi}. By learning distributions of sensor measurements, they could impute plausible values when signals are lost and fuse information across modalities to provide more consistent state estimates. 
For vision-based perception, generative models could also compensate for incomplete or distorted depth information by producing alternative interpretations of the scene \cite{deng2025multi}. 
In this way, MPC could receive a more stable and reliable representation of the system state, enabling coordinated control even when raw sensing and perception are imperfect.

For instance, in robotic grasping tasks where object occlusion degrades depth perception \cite{deng2025multi}, generative models could produce plausible depth maps from partially visible observations and infer object poses when direct 
measurements are unavailable. 
Similarly, in multi-drone systems experiencing 
intermittent localization signals, generative imputation could maintain position estimates during signal loss by fusing data from multiple sensor modalities, providing the MPC with state information necessary for coordinated 
flight control \cite{zhang2022aerial}.

%Addressing Control Design and Tuning Difficulty with Generative ML.
When MPC depends on task-specific tuning of cost weights and parameters, its performance often fails to generalize across different tasks and system configurations \cite{kalaycioglu2023passivity, zhang2022multi}. 
Generative methods could reduce this burden by learning priors over policies and cost weights, offering useful starting points for optimization. 
They could also learn control strategies directly from demonstrations or generate diverse candidate plans that guide MPC, reducing the need for explicit reward specification or extensive manual adjustment \cite{muhammed2024multi, deng2025multi}. 
In this way, MPC could shift from heavy offline tuning toward data-driven initialization and refinement, enabling broader adaptability in multi-robot manufacturing.

For instance, in multi-robot systems where different controller configurations and designs can significantly affect performance \cite{zhang2022multi}. 
Generative models could explore diverse parameter settings and architectural choices to identify optimal control configurations. 
Rather than relying on trial-and-error manual tuning for each scenario, generative priors could provide adaptive initializations that reduce the tuning burden by offering reasonable starting points based on learned patterns from previous configurations. 
This exploration capability could enable MPC to systematically discover effective control settings across varying task conditions and robot configurations.

\begin{figure*}[thbp]
    \centering
    \includegraphics[width=0.9\textwidth]{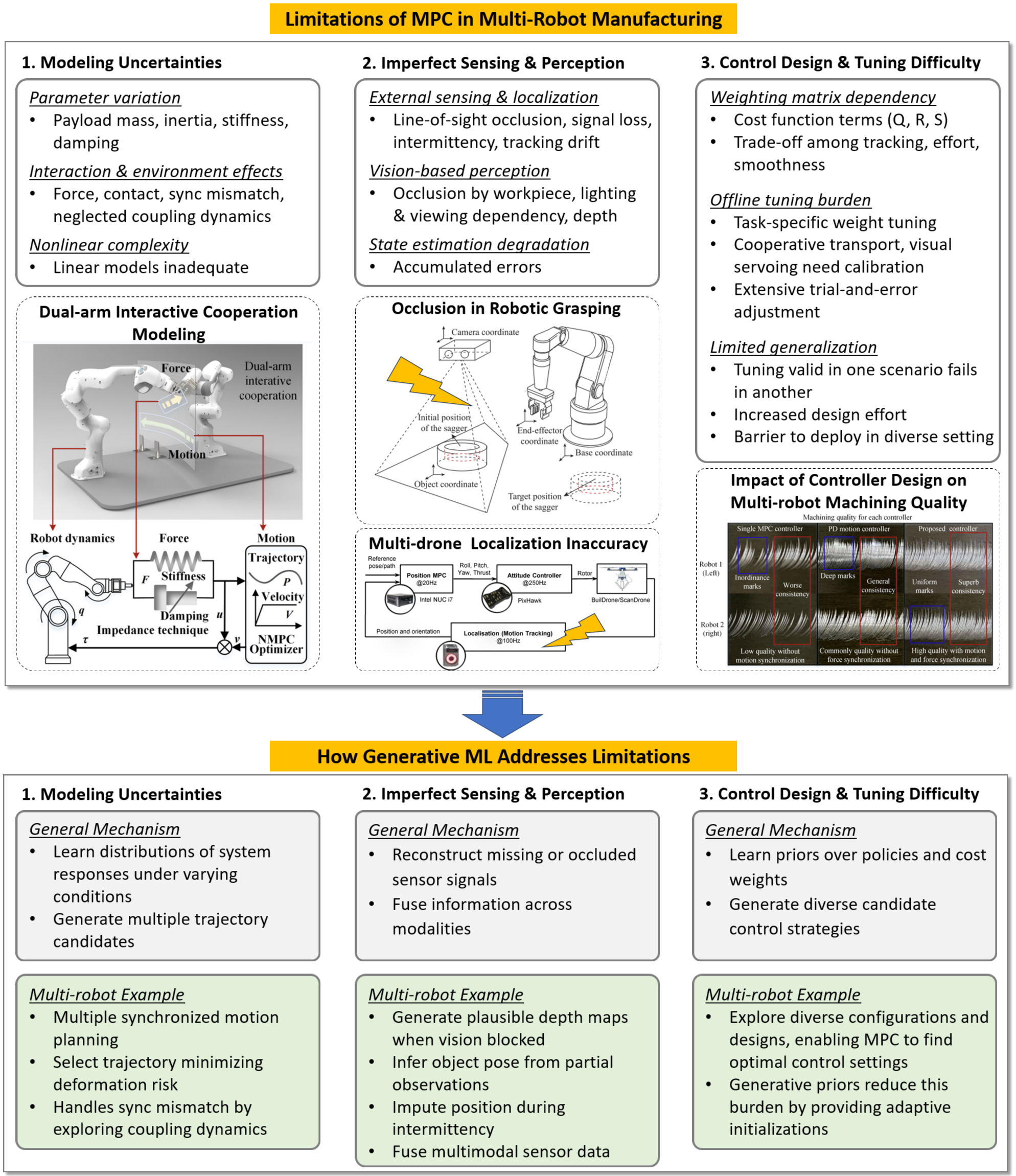}
    \caption{Illustrative case of multi-robot manufacturing showing how generative ML addresses the main limitations of MPC identified through literature review.
    (Top) Three main limitation categories are illustrated: modeling uncertainties from parameter variations and interaction dynamics, shown through dual-arm cooperative operation \cite{zhang2022multi}.
    Imperfect sensing and perception, including occlusion and localization errors, exemplified by robotic grasping \cite{deng2025multi} and multi-drone tracking \cite{zhang2022aerial}.
    Comparative machining outcomes show how controller design and configuration significantly affect quality, underscoring the difficulty of systematic tuning \cite{zhang2022multi}.
    (Bottom) Generative ML mechanisms address these through distribution learning, generative imputation, and learned priors.    
    Although multi-robot systems are used as an example, the concept reflects broader applicability across diverse manufacturing contexts.}

    \label{fig:genml_multi_robot}
\end{figure*}

%Summary.
The discussion above illustrates how generative ML can address the main limitations of MPC in the representative case of multi-robot manufacturing, as shown in Figure \ref{fig:genml_multi_robot}.
While this example was chosen to demonstrate the potential of generative ML for MPC, the same principles can also be applied in other manufacturing contexts. 
Thus, although multi-robot systems serve here as an illustrative case, the integration of generative ML with MPC points more broadly toward a pathway for overcoming similar limitations across diverse applications.

%%%%% Figure %%%%%

%%%%%%%%%%%%%%%%%%

\section{Concluding Remarks}
%Concluding Remarks.
This review has demonstrated the potential of generative ML technologies to enhance MPC for manufacturing processes. 
We surveyed key generative ML methods and their integration with MPC, emphasizing their unique ability to capture uncertainty and provide high-fidelity surrogates that align with advanced manufacturing requirements.
In doing so, we identified critical gaps that still prevent reliable deployment, motivating the need for new research directions.
Taken together, the discussion highlights that generative ML is not simply an incremental add-on to MPC but a means to reshape control for manufacturing. 
By enabling data-driven representations of uncertainty, producing more reliable state information, and supporting control strategies in MPC, generative ML-driven MPC moves manufacturing systems closer to robust and flexible autonomy. 
The multi-robot example illustrates how these benefits can materialize in practice; however, the implications extend broadly to other manufacturing domains facing similar challenges. 
Ultimately, the integration of generative ML with MPC offers a pathway toward next-generation manufacturing control systems that are resilient to uncertainty, adaptable across tasks, and capable of sustaining performance under real-world complexity.

%%%%%%%%%%%%%%%%%%%%%%%%%%%%%%%%%%%%%%%%%%%%%%%%%%%%%%%%%%%%%%%%%%%%%%
\section*{Acknowledgment} %% ASME requests this exact spelling, singular.
The authors would like to acknowledge the support provided by Arizona State University’s faculty startup fund and Fulton Fellowship.

%%%%%%%%%%%%%%%%%%%%%%%%%%%%%%%%%%%%%%%%%%%%%%%%%%%%%%%%%%%%%%%%%%%%%%
%\section*{Funding Data}
%The authors would like to acknowledge the support provided by Arizona State University’s faculty startup fund and Fulton Fellowship.

%%%%%%%%%%%%%
\section*{AI Declaration}
During the preparation of this work, the author(s) used ChatGPT (OpenAI) in order to improve language, clarity, and readability. 
After using this tool, the author(s) reviewed and edited the content as needed and take(s) full responsibility for the content of the publication.

%%%%%%%%%%%%%%%  APPENDICES  %%%%%%%%%%%%%%%%%%%%%%%%%%%%%%%%%%%%%%%%%

%% Note that appendices will be "numbered" A, B, C, ... etc. Use \section, not \section*
%% The equation counter is automatically reset for each appendix
%% Figures will be numbered consecutively with the paper.

%%\appendix % starts appendices

%%%%%%%%%%%%%%%%%%%%%%%%%%%%%%%%%%%%%%%%%%%%%%%%%%%%%%%%%%%%%%%%%%%%%%

%%%%%%%%%%%%%  BIBLIOGRAPHY  %%%%%%%%%%%%%%%%%%%%%%%%%%%%%%%%%%%%%%%%%

%\nocite{*} %% <=== Delete this line - unless you wish to typeset the entire contents of your .bib file.

\bibliographystyle{asmejour}   %% .bst file that follows ASME journal format. Do not change.

%\bibliography{bibliography} %% <=== change this to name of your bib file

%%%%%%%%%%%%%%%%%%%%%%%%%%%%%%%%%%%%%%%%%%%%%%%%%%%%%%%%%%%%%%%%%%%%%%

%% To omit final list of figures and tables, use the class option [nolists]

\end{document}